\documentclass[runningheads]{llncs}
\usepackage[T1]{fontenc}
\usepackage{graphicx}
\usepackage{subfiles}
\usepackage{mathtools}
\usepackage{xspace}
\usepackage[noend]{algpseudocode}
\usepackage{algorithm,algorithmicx}
\usepackage{todonotes}
\usepackage{hyperref}
\usepackage{cleveref}
\usepackage{comment}
\usepackage{amssymb}
\usepackage{textgreek}
\usepackage{float}
\usepackage{cleveref}
\usepackage{pifont}
\usepackage{makecell}

\usepackage{wrapfig}
\algrenewcommand\alglinenumber[1]{\scriptsize #1:}
\begin{document}

\newtheorem{observation}{Observation}
\newcommand{\LB}[1]{\textcolor{blue}{[LB]: #1}}
\newcommand{\ES}[1]{\textcolor[rgb]{0.0,0.0,1.00}{[ES]: #1}}
\newcommand{\MF}[1]{\textcolor[rgb]{1.00,0.00,0.00}{[MF]: #1}}
\newcommand{\TB}[1]{\textcolor[rgb]{0.00,0.50,0.50}{[TB]: #1}}

\newcommand{\rew}[1]{\textcolor{red}{#1}}

\newcommand{\alggoto}{\algorithmicensure{\textbf{goto}\xspace}}
\newcommand*\Let[2]{\State #1 $\gets$ #2}
\algrenewcommand\algorithmicindent{0.3cm}%
\algrenewcommand\algorithmicrequire{\textbf{Precondition:}}
\algrenewcommand\algorithmicensure{\textbf{Postcondition:}}
\algnewcommand{\algorithmicgoto}{\textbf{go to}}%
\algnewcommand{\Goto}[1]{\algorithmicgoto~\ref{#1}}%


\newcommand{\domain}{\ensuremath{\Delta}}
\newcommand{\constantsD}{\ensuremath{\domain_{C}}\xspace}
\newcommand{\nullsD}{\ensuremath{\domain_{N}}\xspace}
\newcommand{\variablesD}{\ensuremath{\domain_{V}}\xspace}
\newcommand{\universalD}{\ensuremath{\domain_{\forall}}\xspace}
\newcommand{\existentialD}{\ensuremath{\domain_{\exists}}\xspace}
\newcommand{\term}{\ensuremath{t}\xspace}
\newcommand{\nulll}{\ensuremath{\varphi}\xspace}
\newcommand{\drule}{\ensuremath{r}\xspace}
\newcommand{\prog}{\ensuremath{P}\xspace}
\newcommand{\setrules}{\ensuremath{\Sigma}\xspace}
\newcommand{\dataset}{\ensuremath{D}\xspace}
\newcommand{\query}{\ensuremath{q}\xspace}
\newcommand{\model}{\ensuremath{M}\xspace}
\newcommand{\terms}{\ensuremath{\mathsf{terms}}}
\newcommand{\vars}{\ensuremath{\mathsf{vars}}}
\newcommand{\const}{\ensuremath{\mathsf{const}}}
\newcommand{\nulls}{\ensuremath{\mathsf{nulls}}}
\newcommand{\fr}{\ensuremath{\mathsf{frontier}}}
\renewcommand{\O}{\ensuremath{\mathcal{O}}}

\newcommand{\PredSymbols}{\ensuremath{\Pi}\xspace}
\newcommand{\structure}{\ensuremath{\varsigma}\xspace}
\newcommand{\mapping}{\ensuremath{\mu}\xspace}
\newcommand{\subst}{\ensuremath{\sigma}\xspace}
\newcommand{\h}{\ensuremath{h}\xspace}
\newcommand{\iso}{\ensuremath{\iota}\xspace}
\newcommand{\arity}{\ensuremath{\mathsf{arity}}}
\newcommand{\maxarity}{\ensuremath{\mathsf{arity}}}
\newcommand{\pos}{\ensuremath{\mathsf{pos}}\xspace}
\newcommand{\apos}{\ensuremath{\pi}\xspace}
\newcommand{\base}{\ensuremath{\mathsf{base}}}
\newcommand{\pred}{\ensuremath{\mathsf{pred}}}
\newcommand{\preds}{\ensuremath{\mathsf{preds}}}
\newcommand{\data}{\ensuremath{\mathsf{data}}}
\newcommand{\rules}{\ensuremath{\mathsf{rules}}}
\newcommand{\affected}{\ensuremath{\mathsf{affected}}}
\newcommand{\unaffected}{\ensuremath{\mathsf{unaffected}}}
\newcommand{\dangerous}{\ensuremath{\mathsf{dangerous}}}
\newcommand{\harmful}{\ensuremath{\mathsf{harmful}}\xspace}
\newcommand{\atoms}{\ensuremath{\mathsf{atoms}}}
\newcommand{\mods}{\ensuremath{\mathsf{mods}}}
\newcommand{\body}{\ensuremath{\mathsf{body}}}
\newcommand{\head}{\ensuremath{\mathsf{head}}}
\newcommand{\hp}{\ensuremath{\mathsf{hp}}}
\newcommand{\bp}{\ensuremath{\mathsf{bp}}}
\newcommand{\dom}{\ensuremath{\mathsf{dom}}}
\newcommand{\idb}{\ensuremath{\mathsf{idb}}}
\newcommand{\edb}{\ensuremath{\mathsf{edb}}}
\newcommand{\cert}{\ensuremath{\mathsf{cert}}}
\newcommand{\ans}{\ensuremath{\mathsf{ans}}}
\newcommand{\ansP}{\ans_\prog}
\newcommand{\fire}{\ensuremath{\mathsf{fire}}}
\newcommand{\isEmpty}{\ensuremath{\mathsf{isEmpty}}}
\newcommand{\update}{\ensuremath{\mathsf{update}}}
\newcommand{\length}{\ensuremath{\mathsf{length}}}
\newcommand{\equals}{\ensuremath{\mathsf{equals}}}
\newcommand{\get}{\ensuremath{\mathsf{get}}}
\newcommand{\size}{\ensuremath{\mathsf{size}}}
\newcommand{\contains}{\ensuremath{\mathsf{contains}}}
\newcommand{\getKey}{\ensuremath{\mathsf{getKey}}}
\newcommand{\add}{\ensuremath{\mathsf{add}}}
\newcommand{\filter}{\ensuremath{\mathsf{filter}}}
\DeclarePairedDelimiter\freeze{\lceil}{\rfloor}
\newcommand{\firingcondition}{\textsc{FiringCondition}\xspace}

\newcommand{\cmark}{\ding{51}}%
\newcommand{\xmark}{\ding{55}}%

\newcommand{\makepred}[1]{\ensuremath{\normalfont{\texttt{#1}}}\xspace}
\newcommand{\makeconstant}[1]{\ensuremath{\normalfont{\texttt{#1}}}\xspace}

\newcommand{\ca}{\makeconstant{a}}
\newcommand{\cb}{\makeconstant{b}}
\newcommand{\cc}{\makeconstant{c}}
\newcommand{\cd}{\makeconstant{d}}
\newcommand{\ce}{\makeconstant{e}}
\newcommand{\cp}{\makeconstant{p}}

\newcommand{\pp}{\makeconstant{p}}
\newcommand{\pq}{\makeconstant{q}}

\newcommand{\makeatom}[1]{\ensuremath{\mathbf{#1}}\xspace}
\newcommand{\maketuple}[1]{\ensuremath{\bar{#1}}\xspace}

\newcommand{\atoma}{\makeatom{a}}
\newcommand{\atomb}{\makeatom{b}}
\newcommand{\atomc}{\makeatom{c}}

\newcommand{\anull}{\ensuremath{\varphi}}

\newcommand{\makevar}[1]{\ensuremath{\texttt{#1}}\xspace}
\newcommand{\makevarn}[2]{\ensuremath{\texttt{#1}_\texttt{#2}}\xspace}
\newcommand{\vx}{\makevar{X}}
\newcommand{\vy}{\makevar{Y}}
\newcommand{\vz}{\makevar{Z}}
\newcommand{\vw}{\makevar{W}}
\newcommand{\vu}{\makevar{U}}
\newcommand{\vnx}[1]{\makevarn{X}{#1}}
\newcommand{\vny}[1]{\makevarn{Y}{#1}}
\newcommand{\vnz}[1]{\makevarn{Z}{#1}}
\newcommand{\vnw}[1]{\makevarn{W}{#1}}
\newcommand{\vnu}[1]{\makevarn{U}{#1}}

\newcommand{\makevarset}[1]{\ensuremath{\mathbf{#1}}\xspace}
\newcommand{\vX}{\makevarset{X}}
\newcommand{\vY}{\makevarset{Y}}
\newcommand{\vZ}{\makevarset{Z}}
\newcommand{\vW}{\makevarset{W}}
\newcommand{\vU}{\makevarset{U}}

\newcommand{\datalogE}{\normalfont{Datalog}\ensuremath{^\exists}\xspace}
\newcommand{\plusminus}{\ensuremath{^\pm}\xspace}
\newcommand{\datalogpm}{\normalfont{Datalog}\plusminus}
\newcommand{\shy}{\ensuremath{\mathsf{shy}}\xspace}
\newcommand{\warded}{\ensuremath{\mathsf{warded}}\xspace}
\newcommand{\harmless}{\ensuremath{\mathsf{harmless}}\xspace}
\newcommand{\protectedd}{\ensuremath{\mathsf{protected}}\xspace}
\newcommand{\wardplus}{\ensuremath{\mathsf{ward}^+}\xspace}
\newcommand{\dyadic}[1]{\ensuremath{\mathsf{dyadic}\text{-}#1}\xspace}
\newcommand{\dyadicshy}{\dyadic{\shy}}
\newcommand{\headground}{\ensuremath{_{\textsc{HG}}}}
\newcommand{\dyadicdecomposition}{\ensuremath{\mathsf{dyadic}\text{-}\mathsf{decomposition}}}

\newcommand{\chase}{\ensuremath{chase}\xspace}
\newcommand{\chaser}{\ensuremath{chase_r}\xspace}
\newcommand{\chases}{\ensuremath{chase_s}\xspace}
\newcommand{\rchase}{\ensuremath{rchase}\xspace}
\newcommand{\ochase}{\ensuremath{ochase}\xspace}
\newcommand{\ichase}{\ensuremath{ichase}\xspace}
\newcommand{\ichaser}{\ensuremath{ichase_r}\xspace}
\newcommand{\ichases}{\ensuremath{ichase_s}\xspace}
\newcommand{\pchase}{\ensuremath{pchase}\xspace}
\newcommand{\pchaser}{\ensuremath{pchase_r}\xspace}
\newcommand{\pchases}{\ensuremath{pchase_s}\xspace}

\newcommand{\chasehom}{\textsc{ChaseHom}\xspace}
\newcommand{\chaseisom}{\textsc{ChaseIsom}\xspace}

\newcommand{\isomorphismcheck}{\textsc{IsomorphismCheck}\xspace}
\newcommand{\homomorphismcheck}{\textsc{HomomorphismCheck}\xspace}
\newcommand{\isomorphismcheckstreaming}{\textsc{IsomorphismCheck$_S$}\xspace}
\newcommand{\homomorphismcheckstreaming}{\textsc{HomomorphismCheck$_S$}\xspace}

\newcommand{\bcqa}{\textsc{bcqa}\xspace}
\newcommand{\abcqa}{\textsc{abcqa}\xspace}

\newcommand{\np}{\textsc{np}\xspace}
\newcommand{\npcomplete}{\np\text{-complete}\xspace}
\newcommand{\ac}{\textsc{AC}\textsubscript{0}\xspace}
\newcommand{\ptime}{\textsc{PTime}\xspace}
\newcommand{\exptime}{\textsc{ExpTime}\xspace}
\newcommand{\dtime}{\textsc{dtime}\xspace}
\newcommand{\logcfl}{\textsc{logcfl}\xspace}
\newcommand{\logcflcomplete}{\logcfl\text{-complete}\xspace}
\newcommand{\ptimecomplete}{\textsc{PTime}\text{-complete}\xspace}

\newcommand{\wardedbcqeval}{\textsc{WardedBCQEval}\xspace}
\newcommand{\ms}{\textsc{MS}\xspace}
\newcommand{\msshy}{\ensuremath{\textsc{MS}_\shy}\xspace}
\newcommand{\streamichase}{\ensuremath{stream}\text{-}\ensuremath{ichase}\xspace}
\newcommand{\lazypchase}{\ensuremath{lazy}\text{-}\ensuremath{pchase}\xspace}
\newcommand{\generated}{\ensuremath{gen}\xspace}
\newcommand{\canonical}{\mathsf{canonical}\xspace}
\newcommand{\hashset}{\ensuremath{\mathit{hashSet}}}
\newcommand{\hashmap}{\ensuremath{\mathsf{hashMap}}}
\newcommand{\proj}{\ensuremath{\mathsf{proj}}\xspace}
\newcommand{\constcanonical}{\ensuremath{\mathsf{constcanonical}}\xspace}
\newcommand{\completedb}{\ensuremath{\mathcal{\dataset}}\xspace}
\newcommand{\vadalog}{\textsc{Vadalog}\xspace}
\newcommand{\dlv}{\textsc{DLV}\xspace}
\newcommand{\dlve}{\textsc{DLV}\ensuremath{^\exists}\xspace}

\newcommand{\vadalogI}{\textsc{Vadalog-i}\xspace}
\newcommand{\vadalogP}{\textsc{Vadalog-p}\xspace}
\newcommand{\vadalogIR}{\textsc{Vadalog-ir}\xspace}
\newcommand{\vadalogPR}{\textsc{Vadalog-pr}\xspace}
\newcommand{\rdfox}{\textsc{RDFox}\xspace}
\newcommand{\llunatic}{\textsc{Llunatic}\xspace}

\newcommand{\unfold}{\ensuremath{\mathsf{unfold}}\xspace}
\newcommand{\fold}{\ensuremath{\mathsf{fold}}\xspace}
\newcommand{\sk}[2]{f_{#1, #2}}

\newcommand{\ei}{\ensuremath{(\mathit{i})}\xspace}
\newcommand{\eii}{\ensuremath{(\mathit{ii})}\xspace}
\newcommand{\eiii}{\ensuremath{(\mathit{iii})}\xspace}
\newcommand{\eiv}{\ensuremath{(\mathit{iv})}\xspace}
\mathchardef\mhyphen="2D 

\newcommand{\outedges}{\ensuremath{\mathsf{out}\text{-}\mathsf{edges}}}
\newcommand{\labell}{\ensuremath{\mathsf{label}}}
\newcommand{\VisitNode}{\textsc{VisitNode}}
%
%
%
\title{Ontological Reasoning over Shy and Warded Datalog+/-- for Streaming-based Architectures (technical report)} 
\titlerunning{Ontological Reasoning for Streaming-based Architectures}
%
 \author{Teodoro Baldazzi\inst{1} \and
 Luigi Bellomarini\inst{2} \and
 Marco Favorito\inst{2} \and\\
 Emanuel Sallinger\inst{3,4}}
%
%
 \institute{Universit\`a Roma Tre, Italy \email{teodoro.baldazzi@uniroma3.it} \and
 Banca d'Italia, Italy \and
 TU Wien, Austria \and
 University of Oxford, UK}
\maketitle              
\begin{abstract}
Recent years witnessed a rising interest towards Datalog-based ontological reasoning systems, both in academia and industry. These systems adopt languages, often shared under the collective name of Data\-log$^\pm$, that extend Datalog with the essential feature of existential quantification, while introducing syntactic limitations to sustain reasoning decidability and achieve a good trade-off between expressive power and computational complexity. From an implementation perspective, modern reasoners borrow the vast experience of the database community in developing streaming-based data processing systems, such as volcano-iterator architectures, that sustain a limited memory footprint and good scalability. In this paper, we focus on two extremely promising, expressive, and tractable languages, namely, Shy and Warded Datalog$^\pm$. We leverage their theoretical underpinnings to introduce novel reasoning techniques, technically, ``chase variants'', that are particularly fit for efficient reasoning in streaming-based architectures. We then implement them in Vadalog, our reference streaming-based engine, to efficiently solve ontological reasoning tasks over real-world settings.
\keywords{Ontological reasoning \and Datalog \and Chase \and Vadalog.}
\end{abstract}
\section{Introduction}
\label{sec:introduction}
In the last decade there has been a growing interest, both academic and industrial, towards novel solutions to perform complex reasoning tasks in an efficient and scalable fashion.
This fostered the development of modern \textit{intelligent systems} enriched with \textit{reasoning capabilities} that employ powerful languages for \textit{knowledge representation}~\cite{GottlobP15,KrotzschT16}.
To enable the full potential of ontological reasoning and effectively address relevant tasks of \textit{ontology-based query answering} (QA) and \textit{knowledge graph} navigation, the languages implemented by such reasoners should provide high expressive power, jointly supporting, for instance, recursion and existential quantification.
At the same time, they should fulfil decidability and tractability requirements for QA, so as to fit large-scale applications~\cite{BellomariniGPS17}.

Among the languages for \textit{Knowledge Representation and Reasoning} (KRR), the \textit{Datalog$^\pm$}~\cite{CaliGLMP10} family became the subject of flourishing research and applications in recent years.
Its members (technically, \textit{fragments}) extend Datalog~\cite{AbiteboulHV95} with existential quantification in rule heads, relying on syntactical restrictions such as \textit{guardedness}~\cite{CaliGK13}, \textit{weak-acyclicity}~\cite{FaginKMP05}, and \textit{stickiness}~\cite{CaliGP10}, etc.,
to prevent the QA undecidability caused by a naive integration of existentials~\cite{CaliGK13}.
Their semantics can be specified via the \textit{chase}~\cite{MaierMS79} procedure, an algorithmic tool that takes as input a database $\dataset$ and a set $\setrules$ of rules, and modifies $\dataset$ by adding new facts until $\setrules$ is satisfied.
In this paper, we consider two particularly promising fragments, namely, \textit{Shy}~\cite{LeoneMTV19} and \textit{Warded}~\cite{CaliGP10,GottlobP15}.
They both cover the requirements for KRR, offering a very good trade-off between expressive power, being able to express all SPARQL queries under OWL 2 QL entailment regime and set semantics, and computational complexity, featuring $\textsf{PTIME}$ in data complexity for Boolean conjunctive QA.
Reasoning over Shy and Warded is based on powerful chase variants that limit the generation of unavailing facts and ensure termination, namely, the \textit{parsimonious} chase, which does not allow creating new facts for which a homomorphism exists to a fact already in $\dataset$, and the \textit{isomorphic} chase, preventing the generation of isomorphic facts instead.

Shy and Warded, together with such chase methodologies, are both implemented in state-of-the-art reasoners, namely \textit{DLV$^\exists$}~\cite{dlvE} and the \textit{Vadalog}~\cite{BellomariniSG18,BellomariniBGS22} system.
While the data complexity of Shy and Warded is the same, their representative reasoners work quite differently in practice.
In fact, $\dlve$, which extends $\dlv$~\cite{LeonePFEGPS06} to reason over Shy, employs a \textit{materialization} approach, producing and storing all the facts via \textit{semi-na\"ive} evaluation~\cite{AbiteboulHV95}, in which the rules in $\Sigma$ are activated over $\dataset$ according to a bottom-up \textit{push-based} strategy.
It requires reading the whole $\dataset$ first and then keeping both the original and the newly generated facts in memory throughout the entire reasoning, which is impractical with large and complex datasets~\cite{Pitoura2018}.
Conversely, to enable scalable QA with a limited memory footprint, recent reasoners like Vadalog leverage the vast experience of the database community and adopt \textit{streaming-based} data processing architectures based, for example, on the \textit{volcano-iterator model}~\cite{GraefeM93}.
They operate in a \textit{pull-based} \textit{query-driven} fashion in which, ideally, facts are materialized only at the end of the evaluation and if they contribute to the QA~\cite{BellomariniBGS22}.

The chase is naturally fit for bottom-up evaluations and materialization, yet the need for repeated homomorphism checks to avoid generating duplicates and the consequential requirement to keep in memory an updated version of $\dataset$, or at least stratum-based intermediate results, highly affects scalability.
At the same time, the streaming architectures, inherently memory-bound, are naturally incompatible with these computational tactics.
Enabling powerful chase variants such as isomorphic and parsimonious, which tend to be significantly smaller than semi-na\"ive chase due to the more restrictive homomorphism condition~\cite{BaldazziBFS22}, would allow to better exploit the scalability of the streaming architecture for QA.

\smallskip \noindent
\textbf{Contribution.}
This paper strives to leverage the theoretical underpinnings of Shy and Warded and propose for the first time, to the best of our knowledge, novel reasoning methodologies that are highly suited for streaming-based architectures.
Bridging the gap between the theory and practice of these two fragments, we develop an effective \textit{streaming-friendly} chase variant for streaming-based reasoning engines, implementing and evaluating it in the Vadalog system.
More in detail, in this paper we provide the following contributions.
\begin{itemize}\setlength\itemsep{0.2em}
    \item We \textbf{systematize the semantic relationship} between Shy and Warded w.r.t. the applicability of parsimonious and isomorphic chase for QA.
    \item We \textbf{develop novel variants of isomorphic and parsimonious chase}, suitable in a streaming-based environment for ontological reasoning.
    \item We \textbf{discuss the integration of the chase procedure in Vadalog streaming architecture}, presenting novel foundations and features of the system.
    \item We \textbf{provide an experimental comparison of Vadalog} with modern materialization-based systems, showing its efficiency over complex QA tasks.
\end{itemize}

\noindent
\textbf{Related Work.}
Modern reasoners that encapsulate Datalog$^\pm$ expressive power and make use of chase-based methodologies recently emerged in the literature.
Relevant examples include, but are not limited to, the already mentioned \textit{DLV$^\exists$} \cite{LeoneMTV19} with its parsimonious chase, and \textit{DLV}~\cite{LeonePFEGPS06}, 
\textit{Llunatic}~\cite{GeertsMPS14}, and \textit{RDFox}~\cite{NenovPMHWB15} based on the \textit{Skolem} chase instead~\cite{MeccaPR12}.
While featuring different reasoning capabilities and performance, they share the common trait of a materialization-based architecture that employs variants of the semi-na\"ive evaluation.
Vadalog is, to our knowledge, the first fully-fledged Datalog$^\pm$ reasoner that effectively synergizes powerful chase methodologies with a streaming-based architecture, with very good results in performance and scalability over real-world tasks.

\smallskip \noindent
\textbf{Overview.}
In Section~\ref{sec:section2} we discuss chase-based reasoning over Shy and Warded.
In Section~\ref{sec:section3} we present the streaming-friendly versions of the chase.
In Section~\ref{sec:section4} we illustrate the chase integration in the Vadalog architecture.
Section~\ref{sec:section5} features the experimental evaluation and we draw our conclusions in Section~\ref{sec:conclusion}.
The Appendix delves into the theoretical results and corresponding proofs.
\section{Chase-based Reasoning over Shy and Warded}
\label{sec:section2}
To guide our discussion, we provide relevant results on ontological reasoning over Shy and Warded with the chase, also taking into account their recently introduced intersection Protected~\cite{BaldazziBFS22}.

\subsection{Preliminaries}
Let $\constantsD$, $\nullsD$, and $\variablesD$ be disjoint countably infinite sets of \textit{constants}, \textit{nulls} and \textit{variables}, respectively.
A (\textit{relational}) \textit{schema} $P$ is a finite set of relation symbols (or \textit{predicates}) with associated arity.
A \textit{term} is either a constant or a variable.
An \textit{atom} over $P$ is an expression of the form $R(\bar v)$, where $R \in P$ is of arity $n > 0$ and $\bar v$ is an $n$-tuple of terms. 
A \textit{database} (\textit{instance}) over $P$ associates to each symbol in $P$ a relation of the respective arity over the domain of constants and nulls.
The members of the relations are called \textit{tuples} or \textit{facts}.
Given a fact $\atoma$, $\const(\atoma)$ is the set of constants and $\nulls(\atoma)$ is the set of nulls in $\atoma$.
Given two conjunctions of atoms \textvarsigma$_1$ and \textvarsigma$_2$, a \textit{homomorphism} from \textvarsigma$_1$ to \textvarsigma$_2$ is a mapping $h : \constantsD \cup \nullsD \cup \variablesD \to \constantsD \cup \nullsD \cup \variablesD$ s.t. $h(t)=t$ if $t \in \constantsD$, $h(t) \in \constantsD \cup \nullsD$ if $t \in \nullsD$ and for each atom $a(t_1,\ldots,t_n) \in$ \textvarsigma$_1$, then $h(a(t_1,\ldots,t_n))=a(h(t_1),\ldots,h(t_n)) \in$ \textvarsigma$_2$.
An \textit{isomorphism} between \textvarsigma$_1$ and \textvarsigma$_2$ is a homomorphism $h$ from \textvarsigma$_1$ to \textvarsigma$_2$ s.t. $h^{-1}$ is a homomorphism from \textvarsigma$_2$ to \textvarsigma$_1$.

\smallskip \noindent \textbf{Syntax and Dependencies.}
A Datalog$^\pm$ program $\Pi$ consists of a set of tuples and \textit{tuple-generating dependencies} (TGDs), i.e., function-free Horn clauses of the form $\forall \bar x \forall \bar y (\varphi(\bar x,\bar y)$$\to$$\exists \bar z~\psi(\bar x,\bar z))$, where $\varphi(\bar x,\bar y)$ (the \textit{body}) and $\psi(\bar x,\bar z)$ (the \textit{head}) are conjunctions of atoms, $\bar x, \bar y$ are vectors of universally quantified variables (\textit{$\forall$-variables}) and constants, and $\bar z$ is a vector of existentially quantified variables (\textit{$\exists$-variables}).
Quantifiers can be omitted and conjunction is denoted by comma.
Given a set $\Sigma$ of Datalog$^\pm$ rules and a position $R[i]$ (i.e., the $i$-th term of a predicate $R$ with arity $k$, where $i=1,\ldots,k$), $R[i]$ is \textit{affected} if (i)~$R$ appears in a rule in $\Sigma$ with an \textit{$\exists$-variable} in $i$-th term or, (ii)~there is a rule in $\Sigma$ such that a \textit{$\forall$-variable} is only in affected body positions and in $R[i]$ in the head.

\smallskip \noindent \textbf{Shy, Warded and Protected.}
Let $\Sigma$ be a set of Datalog$^\pm$ rules.
A position $R[i]$ is \textit{invaded} by an \textit{$\exists$-variable} $y$ if there is a rule $\sigma$ $\in$ $\Sigma$ such that \textit{head}($\sigma$) $=$ $R$($t_1,\ldots,t_k$) and either (i)~$t_i=y$ or, (ii)~$t_i$ is a $\forall$-variable that occurs in \textit{body($\sigma$)} only in positions invaded by $y$.
Thus, if $R[i]$ is invaded, then it is affected, but not vice versa.
Let $x$ $\in$ $\textbf{X}$ be a variable in a conjunction of atoms \textvarsigma$_{[\textbf{X}]}$.
Then $x$ is \textit{attacked} in \textvarsigma$_{[\textbf{X}]}$ by $y$ if $x$ occurs in \textvarsigma$_{[\textbf{X}]}$ only in positions invaded by $y$.
If $x$ is not attacked, it is \textit{protected} in \textvarsigma$_{[\textbf{X}]}$.
Thus, $\Sigma$ is \textit{Shy} if, for each rule $\sigma$ $\in$ $\Sigma$: (i)~if a variable $x$ occurs in more than one body atom, then $x$ is protected in \textit{body}($\sigma$); and, (ii)~if two distinct $\forall$-variables are not protected in \textit{body}($\sigma$) but occur both in \textit{head}($\sigma$) and in two different body atoms, then they are not attacked by the same variable~\cite{LeoneMTV19}.
A $\forall$-variable $x$ is \textit{harmful}, wrt a rule $\sigma$ in $\Sigma$, if $x$ appears only in affected positions in $\sigma$, otherwise it is \textit{harmless}.
A (join) rule that contains a harmful (join) variable is a \textit{harmful} (\textit{join}) \textit{rule}.
If the harmful variable is in \textit{head($\sigma$)}, it is \textit{dangerous}.
Thus, $\Sigma$ is \textit{Warded} if, for each rule $\sigma$ $\in$ $\Sigma$: (i)~all the dangerous variables appear in a single body atom, called \textit{ward}; and, (ii)~the ward only shares harmless variables with other atoms in the body~\cite{GottlobP15}.
Without loss of generality (as more complex joins can be broken into steps~\cite{BellomariniBGS22}), an \textit{attacked harmful join} rule $\tau$: $A(x_1,y_1,h),B(x_2,y_2,h) \to \exists{z}~C(\overline{x},z)$ is a rule in $\Sigma$ where $A[3]$ and $B[3]$ are positions invaded by (at least) one common $\exists$-variable, $x_1, x_2 \subseteq \overline{x}$, $y_1, y_2 \subseteq \overline{y}$ are disjoint tuples of harmless variables or constants and $h$ is an attacked harmful variable.
If $h$ is otherwise protected, $\tau$ is a \textit{protected harmful join} rule.
Thus, $\Sigma$ is \textit{Protected} if, for each rule $\sigma$ $\in$ $\Sigma$: (i)~$\sigma$ does not contain attacked harmful joins; and, (ii)~$\sigma$ is Warded.
Protected corresponds to the \textit{intersection} between Shy and Warded~\cite{BaldazziBFS22}.
The rewriting of a Warded set $\Sigma$ of rules with attacked harmful joins into an equivalent
Protected set is achieved via the \textit{Attacked Harmful Join Elimination} (AHJE).
It replaces each attacked harmful join rule $\tau$ with a set of protected rules that cover the generation of all the facts derived from activating $\tau$, thus preserving correctness~\cite{BaldazziBSA21}.

\smallskip \noindent \textbf{Reasoning and Query Answering.}
An ontological reasoning task consists in answering a \textit{conjunctive query} (CQ) $Q$ over a database $D$, augmented with a set $\Sigma$ of rules.
More formally, given a database $D$ over $P$ and a set of TGDs $\Sigma$, we denote the \textit{models} of $D$ and $\Sigma$ as the set $\mathbf{B}$ of all databases (and we write $\mathbf{B} \models D \cup \Sigma$) such that $\mathbf{B} \supseteq D$, and $\mathbf{B} \models \Sigma$.
A conjunctive query $Q$ is an implication $q(\bar x)\leftarrow\psi(\bar x,\bar z)$, where $\psi(\bar x,\bar z)$ is a conjunction of atoms over $P$, $q(\bar x)$ is an $n$-ary predicate $\notin$ $P$, and $\bar x,\bar z$ are vectors of variables and constants.
A \textit{Boolean} CQ (BCQ) $Q\leftarrow\psi(\bar x, \bar z)$ over $D$ under $\Sigma$ is a type of CQ whose answer is \textit{true} (denoted by $D\models q$) iff there exists a homomorphism \textit{h}: $\constantsD \cup \variablesD$ $\rightarrow$ $\constantsD \cup \nullsD$ s.t. \textit{h}($\psi(\bar x, \bar z)$) $\subseteq$ $D$.
It is known that the query output tuple problem (i.e., the decision version of CQ evaluation) and BCQ evaluation are AC$_0$-reducible to each other~\cite{CaliGL12}.
Thus, for simplicity of exposition and without loss of generality, we will state our results in terms of BCQ Answering (BCQA).

\smallskip \noindent \textbf{Semantics and Chase.}
The semantics of a Datalog$^\pm$ program can be defined in an operational way with the \textit{chase procedure}~\cite{JohnsonK84,MaierMS79}.
It enforces the satisfaction of a set $\Sigma$ of rules over a database $D$, incrementally expanding $D$ with facts entailed via the application of the rules over $D$, until all of them are \textit{satisfied}.
Such facts possibly contain fresh new symbols $\nu$ (\textit{labelled nulls}) to satisfy existential quantification.
A TGD $\sigma:$ $\varphi(\bar x,\bar y)$$\to$$\psi(\bar x,\bar z)$ is satisfied by $D$ if, whenever a homomorphism $\theta$ occurs (is \textit{fired}) such that $\theta(\varphi(\bar x,\bar y))$ $\subseteq$ $D$, there exists an \textit{extension} $\theta^\prime$ of $\theta$ (i.e., $\theta$ $\subseteq$ $\theta^\prime$) such that $\theta^\prime(\psi(\bar x,\bar z))$ $\subseteq$ $D$.
In the na\"ive chase (namely, \textit{oblivious} or $\ochase$), an applicable homomorphism $\theta$ from $\sigma$ over $D$ occurs if $\theta(\varphi(\bar x,\bar y))$ $\subseteq$ $D$.
When applied, it generates a new fact $\theta^\prime(\psi(\bar x,\bar z))$ that enriches $D$, if not already present, where $\theta^\prime$ extends $\theta$ by mapping the variables of $\bar z$ (if not empty) to new nulls named in a lexicographical order~\cite{CaliGL12}.
Without loss of generality, we assume nulls introduced at each fire functionally depend on the pair $\langle \sigma, \theta\rangle$ that is involved in the fire.
Regardless of the order in which applicable homomorphisms are fired, $\ochase(D,\Sigma)$ is unique.

\subsection{Boolean QA over Shy and Warded with the Chase}
In the joint presence of recursion and existentials, an infinite number of nulls could be generated in $\ochase$, inhibiting termination and QA decidability~\cite{CaliGL12}.
\begin{example}
\textit{Consider the following Datalog$^\pm$ set $\Sigma$ of rules}
\label{ex:running-example}
    \begin{align*}
        \textit{Employee}(x)\rightarrow\exists{s}~\textit{WorksFor}(x,s) \tag{$\alpha$}\\
        \textit{HasBoss}(x,y), \textit{WorksFor}(x,s)\rightarrow\textit{WorksFor}(y,s) \tag{$\beta$}\\
        \textit{WorksFor}(x,s), \textit{WorksFor}(y,s)\rightarrow\textit{Knows}(x,y) \tag{$\gamma$}\\
        \textit{Knows}(x,y)\rightarrow\exists{s}~\textit{WorksFor}(x,s), \textit{WorksFor}(y,s) \tag{$\delta$}
    \end{align*}
    \textit{For each employee $x$ there exists an entity $s$ that $x$ works for (rule~$\alpha$). If $x$ has $y$ as boss, then $y$ also works for $s$ (rule~$\beta$). If $x$ and $y$ work for the same $s$, then they know each other (rule~$\gamma$) and vice-versa (rule~$\delta$). Consider the database $D$ =\{\textit{Employee}(\textit{Alice}), \textit{Employee}(\textit{Bob}), \textit{HasBoss}(\textit{Alice},\textit{Bob})\}.}
\end{example}

\noindent
Due to the existential quantification in rule~$\delta$ and its interplay with the recursion in rules~$\beta$ and~$\gamma$, the result of computing $\ochase(D,\Sigma)$ contains an infinite set $\bigcup_{i=1,\ldots}\{\textit{WorksFor}(\textit{Alice},\nu_i),$ $\textit{WorksFor}(\textit{Bob},\nu_i)\}$.
To cope with this, Datalog$^\pm$ fragments make use of distinct versions of the $\ochase$ based on \textit{firing conditions} to limit the applicability of the homomorphisms and preserve termination and decidability.
Among them, we focus on the \textit{parsimonious} (or $\pchase$) and the \textit{isomorphic} (or $\ichase$) chase.
Consider a database $I^\prime$.
In the former, an applicable homomorphism $\theta$ of a rule $\sigma$ is fired if, additionally, there is no homomorphism from $\theta(\head(\sigma))$ to $I^\prime$ ($\homomorphismcheck$)~\cite{LeoneMTV19}.
In the latter, $\theta$ is fired if, additionally, there is no isomorphic embedding of $\theta(\head(\sigma))$ to $I^\prime$ ($\isomorphismcheck$)~\cite{BellomariniBGS22}.
Observe that for any database $D$, $\datalogpm$ set $\Sigma$ of rules and query $\query$, $\pchase(D,\Sigma)\subseteq\ichase(D,\Sigma)$, since the homomorphism check is a stricter firing condition than the isomorphism one, and that they are finite (\cite[Prop. 3.5]{LeoneMTV19} and~\cite[Thm. 3.14]{BellomariniBBS22}).
We now investigate their applicability
over Shy and Warded.
Note that query answering over them is $\exptime$-complete in combined complexity, and $\ptime$-complete in data complexity.

\smallskip \noindent \textbf{Atomic BQA over Shy and Warded.}
Observe that both $\pchase$ and $\ichase$ proved to work only to answer atomic queries, whereas they do not ensure correctness for generic BCQA~\cite{LeoneMTV19,BellomariniBGS22}.
Now, given a Shy set $\Sigma$ of rules, a database $D$, and a BAQ $\query$, we recall that  $\ochase(D,\Sigma)\models\query \text{ iff } \pchase(D,\Sigma)\models\query$~\cite[Thm. 3.6]{LeoneMTV19}.
Similarly, since $\pchase(D,\Sigma)\subseteq\ichase(D,\Sigma)$, we have that $\ochase(D,\Sigma)\models\query \text{ iff } \ichase(D,\Sigma)\models\query$.
Both claims also hold for Protected, since it is the intersection of Shy and Warded.
On the other hand, neither $\pchase$ nor $\ichase$ are complete for Boolean Atomic Query Answering (BAQA) over Warded.
We can disprove completeness by counterexample via the Warded $\Sigma$ in Example~\ref{ex:running-example} and by considering $\query=\textit{Knows}(\textit{Alice},\textit{Bob})$.
Indeed, there exists a rule activation order in which the result of computing $\ichase(D,\Sigma)$ is $D$ $\cup$ \{\textit{WorksFor}(\textit{Alice},$\nu_1$),\textit{WorksFor} (\textit{Bob},$\nu_2$),\textit{Knows}(\textit{Alice},\textit{Alice}),\textit{Knows}(\textit{Bob},\textit{Bob}\}.
Now, \\ \textit{WorksFor}(\textit{Bob},$\nu_1$) is not added to $\ichase(D,\Sigma)$, as it is isomorphic with \textit{WorksFor}(\textit{Bob},$\nu_2$).
The attacked harmful join rule $\gamma$ does not generate $\textit{Knows}(\textit{Ali}\-\textit{ce}, \textit{Bob})$ and $\ichase(D,\Sigma)\not\models \query$.
The same claim holds for $\pchase$ by definition.

\smallskip \noindent \textbf{Conjunctive BQA over Shy and Warded.}
To cover BCQA decidability and preserve the correctness of the evaluation, the chase can be extended with the \textit{resumption} technique~\cite{LeoneMTV19}.
Originally developed for $\pchase$, it consists in iteratively ``resuming'' the procedure in the same state it was after termination (i.e., when the answer to the BCQ is \textit{true} or all the applicable homomorphisms have been examined), performing a promotion of the labelled nulls to constants by the firing condition.
More formally, we call \textit{freezing} the act of promoting a null from $\nullsD$ to a novel constant in $\constantsD$, and given a database instance $I$ , we denote by $\freeze{I}$ the set obtained from $I$ after freezing all of its nulls.
We denote $\pchaser(D,\Sigma,i)$ the $i$-th iteration of resumption for $\chase$, 
where $\pchase(D,\Sigma,0) = D$ and 
$\pchase(D,\Sigma,i) = \pchase(\freeze{\pchase(D,\Sigma,i-1)}, \Sigma)$ for $i\ge 1$.
Furthermore, we can extend the definition of $\ichase$, to perform BCQA, by introducing the \textit{isomorphic chase with resumption} ($\ichaser$), analogous to the $\pchaser$.
Formally, $\ichaser(D,\Sigma,0) = D$, and $\ichaser(D,\Sigma,i) = \ichase(\freeze{\ichaser(D,\Sigma,i-1)}, \Sigma)$ for $i\ge 1$.
Finiteness follows from the definition of $\pchase$, $\ichase$, and resumption.
Indeed, the maximum number of iterations for the resumption that can be performed to answer a certain BCQ $q$ depends on the query itself, and it corresponds to the number of variables in the BCQ $|\vars(q)| + 1$.
We interchangeably adopt the notation $\chaser(D,\Sigma,|\vars(q)| + 1)$ and $\chaser(D,\Sigma)$.
Now, given a Shy set $\Sigma$ of rules, a database $D$, and a BCQ $\query$, we recall that $\ochase(D,\Sigma)\models\query \text{ iff } \pchaser(D,\Sigma,|\vars(q)| + 1)\models\query$~\cite[Thm. 4.11]{LeoneMTV19}. 
Similarly, since $\pchase(D,\Sigma)\subseteq\ichase(D,\Sigma)$ and by definition of resumption, we have that $\ochase(D,\Sigma)\models\query \text{ iff } \ichaser(D,\Sigma)\models\query$.
Both claims also hold for Protected, since it is the intersection of Shy and Warded.
On the other hand, leveraging the above results for BAQA by counterexample via Example~\ref{ex:running-example}, we conclude that neither $\pchaser$ or $\ichaser$ are complete for BCQA over Warded.
\section{Streaming-friendly Firing Conditions in the Chase}
\label{sec:section3}
Leveraging the theoretical underpinnings provided in Section~\ref{sec:section2}, we now lay a foundation towards the integration of the chase in streaming-based reasoning environments by proposing novel variants of homomorphism- and isomorphism-based firing conditions ($\homomorphismcheckstreaming$ and $\isomorphismcheckstreaming$, resp.) for Shy and Protected Datalog$^\pm$ that avoid fact materialization and strive to achieve low time and memory consumption.
Indeed, streaming architectures involve data processing with low memory footprint, without accessing the whole database or fully materializing intermediate results, thus are naturally incompatible with standard $\pchase$ and $\ichase$.

\subsection{Aggregate-based Homomorphism Check}
We recall that a fact $\atoma$ is homomorphic to a fact $\atomb$ if they belong to the same predicate, $\atomb$ features the same constants as $\atoma$ in the same positions, and there exists a mapping of the labelled nulls in $\atoma$ to the constants and nulls in $\atomb$.
To present a streaming-friendly homomorphism-based (i.e., parsimonious) firing condition ($\homomorphismcheckstreaming$), we first introduce the notion of \textit{aggregate fact tree} in the chase.

\begin{definition}
\label{def:aggregate-fact-tree}
Given a predicate $p$ and a set $I$ of facts, an \emph{aggregate fact tree (af-tree)} $T_p$ is a tree s.t.: (i)~the root $T$ of $T_p$ is labelled by $[~]$; (ii)~a fact $\atoma = p(\bar{a}) = p(a_0,\dots,a_k) \in I$ iff there exists a path of $k$ nodes $n_{\bar{a}_0},\dots,n_{\bar{a}_k}$, where $n_{\bar{a}_i}$ ($0 \leq i \leq k$) is labelled by $[a_0,\dots,a_i]$, in $T_p$; and, (iii)~two nodes $n_{\bar{a}_j}$ and $n_{\bar{a}_{j+1}}$ $\in T_p$, labelled by $[a_0,\dots,a_j]$ and $[a_0,\dots,a_{j+1}]$ ($0\le j < k$), respectively, iff there exists an edge from $n_{\bar{a}_j}$ to $n_{\bar{a}_{j+1}}$ labelled by $a_{j+1}$.
\end{definition}

\noindent
Indeed, the path from the root node $T$ to a leaf $n_{\bar{a}}$ denotes that the fact $p(\bar{a})$ $\in$ $I$.
By construction, facts of a predicate $p$ that share the same arguments up to a position $i$ will share the same root-to-node path up to length $i$ in the aggregate fact tree $T_p$.
Thus, the following statement holds.

\begin{proposition}
    \label{prop:homomorphic-streaming}
    Let $\atoma = p(a_1,\dots,a_k)$ be a fact, $I$ a set of facts s.t. $\atoma \notin I$, and $T_p$ the af-tree for $\pred(\atoma) = p$ in $I$. Then there exists a homomorphism $\theta$ from $\atoma$ to $\atomb \in I$ iff there exists a root-to-leaf path $t$ in $T_p$ s.t. $\theta(\atoma[i]) = t[i]$, $1 \leq i \leq k$.
\end{proposition}

\noindent \textbf{Algorithm Overview.}
Leveraging~\Cref{def:aggregate-fact-tree} and~\Cref{prop:homomorphic-streaming}, we devise a procedure, shown in \Cref{alg:homomorphismcheckstreaming-firing-condition}, to perform a novel firing condition for the parsimonious chase with low memory footprint.

\begin{wrapfigure}[9]{r}{0.35\textwidth}
\centering
\vspace{-1.2cm}
\includegraphics[width=0.35\textwidth]{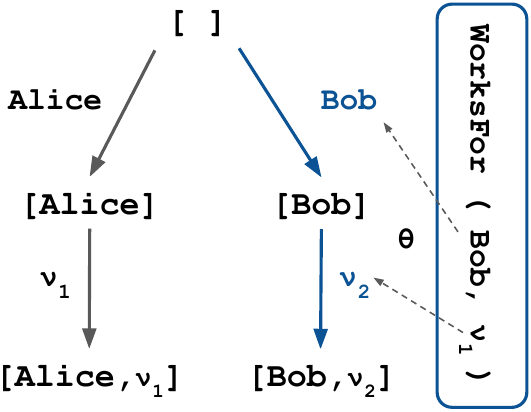}
\caption{T$_\textit{WorksFor}$.}
\label{fig:homomorphismcheckstreaming}
\end{wrapfigure}

\noindent
Intuitively, it verifies whether a new fact $\atoma$ resulting from an applicable homomorphism in the chase is homomorphic to previously generated ones by performing a form of \textit{path matching} in the corresponding aggregate fact tree.
Indeed, if there exists a root-to-leaf path such that a homomorphism occurs from $\atoma$'s arguments to the ones in the path, then the firing condition prevents the applicable homomorphism from being fired and $\atoma$ from being generated.
More specifically, given $\atoma = p(a_1,\dots,a_k)$ and its corresponding aggregate fact tree $T_p$, the algorithm recursively visits the tree in a depth-first fashion via the function \textsc{VisitNode} (line~\ref{alg:line:visit_node}) and checks whether a homomorphism holds from $\atoma$ to the arguments labelling the edges of a root-to-leaf path.
It makes use of a \textit{nullMap} structure to keep track of the mappings from the labelled nulls in $\atoma$ to constants and nulls in the path.

\begin{wrapfigure}[]{L}{0.6\textwidth}
\vspace{-1.5cm}
{\begin{minipage}{\linewidth}
\begin{algorithm}[H]
\scriptsize
    \begin{algorithmic}[1]

    \Function{VisitNode}{$\mathit{currNode}, \mathit{nullMap}, a_i$}
    \If{$\mathit{currNode}$ is a leaf node} 
        \State\Return\ $\textbf{true}$ \label{alg:line:leaf}
    \EndIf
    \If{$a_i$ is a constant} 
        \If{$\nexists$ edge $e=(\mathit{currNode}, v)$ s.t. $\labell(e) = a_i$} \label{alg:line:constant}
            \State \Return \textbf{false}
        \EndIf
        \State \Return $\VisitNode(v,\mathit{nullMap}, t_{i+1})$
    \Else 
        \If{$\exists (a_i,b)\in\mathit{nullMap}$}
            \If{$\nexists$ edge $e =(\mathit{currNode},v) \text{ s.t. } \labell(e) = b$}\label{alg:line:null_mapped}
                \State \Return \textbf{false}
            \EndIf
            \State \Return $\VisitNode(v,\mathit{nullMap},a_{i+1})$
        \EndIf
        \For{edge $e=(\mathit{currNode},v)$} \label{alg:line:null_notmapped_start}
            \Let{$\mathit{nullMap}$}{$\mathit{nullMap}\cup (a_i, \labell(e))$}
            \If{$\VisitNode(v,\mathit{nullMap}, a_{i+1})$}
                \State\Return \textbf{true}
            \EndIf
            \Let{$\mathit{nullMap}$}{$\mathit{nullMap} \setminus \{ (a_i,\labell(e)) \}$} \label{alg:line:null_notmapped_end}
        \EndFor
        \State \Return \textbf{false}
    \EndIf
    \EndFunction
    \State
    \Function{AggrHom\_FiringCondition}{$\atoma$}
        \Let{$T_{\pred(\atoma)}$}{$\mathsf{getAggregateFactTree(\pred(\atoma))}$}
        \State{$\mathit{currNode} \gets T$; $\mathit{nullMap} \gets \emptyset$}
        \If{$\VisitNode(\mathit{currNode},\mathit{nullMap},\atoma[0])$} \label{alg:line:visit_node}
            \State \Return \textbf{false}
        \EndIf
        \State $T_{\pred(\atoma)}.\mathsf{addFact(\atoma)}$
        \State \Return \textbf{true}
    \EndFunction
    \end{algorithmic}
\caption{$\homomorphismcheckstreaming$.}
\label{alg:homomorphismcheckstreaming-firing-condition}
\end{algorithm}
\end{minipage}}
\vspace{-0.8cm}
\end{wrapfigure}

\noindent
Given the current argument $a_i$ and the reached node \textit{currNode}, if the latter is a leaf of the tree, then the homomorphism has been found (line~\ref{alg:line:leaf}).
Otherwise, the homomorphism check over $a_i$ occurs.
If $a_i$ is a constant, then the algorithm checks whether such a term also labels one of the outgoing edges of \textit{currNode} (line~\ref{alg:line:constant}), in which case the target node $v$ is visited.
Otherwise, a homomorphism could not be found and the function returns false.
If $a_i$ is a null instead, distinct behaviours occur depending on whether \textit{nullMap} features a mapping from $a_i$ to an argument $b$ in a previous edge of the path.
If it does, then by definition of mapping function only a node in the tree whose incoming edge is labelled by $b$ can be visited next (line~\ref{alg:line:null_mapped}).
Otherwise, the algorithm carries on the depth-first visit of the tree from \textit{currNode}, attempting to map $a_i$ to the label of each outgoing edge (lines~\ref{alg:line:null_notmapped_start}-\ref{alg:line:null_notmapped_end}).
Finally, if the homomorphism is not found, $T_{\pred(\atoma)}$ is updated with a new path corresponding to $\atoma$'s arguments and the firing condition returns \textit{true}, thus enabling the generation of $\atoma$ in the chase.

The inherent nature of a homomorphism-based approach causes the parsimonious firing condition to be unsuitable for streaming-based architectures.
Nevertheless, employing the aggregate tree data structure and performing the homomorphism checks as a path matching task achieves very good time performance, as shown in Section~\ref{sec:section5}, while limiting memory consumption in the general case.
Figure~\ref{fig:homomorphismcheckstreaming} shows the aggregate fact tree corresponding to predicate \textit{WorksFor} in Example~\ref{ex:running-example}.
Let us consider a new fact \textit{WorksFor}(\textit{Bob},$\nu_1$).
Its generation is prevented by Algorithm~\ref{alg:homomorphismcheckstreaming-firing-condition} due to the existence of path $t$, corresponding to fact \textit{WorksFor}(\textit{Bob},$\nu_2$) in the chase.

\subsection{Hash-based Isomorphism Check}
We recall that a fact $\atoma$ is isomorphic with a fact $\atomb$ if they belong to the same predicate, feature the same constants in the same positions, and there exists a bijection of their labelled nulls.
To present a streaming-friendly isomorphism-based firing condition ($\isomorphismcheckstreaming$), we first introduce the notion of \textit{null canonicalization}.

\begin{definition}
\label{def:null-canonicalization}
Let $\nullsD^c$ be a set of numbered labelled nulls $\{\anull^c_1,\anull^c_2,\dots,\anull^c_n\}$ not appearing in $\nullsD$ or in the set $I$ of facts generated in the chase. 
The \textnormal{canonicalization} of a fact $\atoma$, denoted $\canonical(\atoma)$, is a new fact $\atoma^c$ s.t.: (i)~$\pred(\atoma^c) = \pred(\atoma)$; (ii)~if $\atoma[i] \in \const(\atoma)$, then $\atoma^c[i] = \atoma[i]$; and, (iii)~if $\atoma[i] \in \nulls(\atoma)$, then $\atoma^c[i] = \anull^c_j$, where $j=\arg\min_{k} \atoma[k] = \atoma[i]$. A fact $\atoma$ is \textit{canonical} if $\nulls(\atoma)\subseteq\nullsD^c$.
\end{definition}

\noindent
Intuitively, given a fact $\atoma$, its canonicalization $\atoma^c = \canonical(\atoma)$ differs from it only if $\atoma$ includes labelled nulls, which are replaced by numbered fresh nulls $\anull^c_j$, where $j$ is the index of the first position in $\atoma^c$ featuring $\atoma[i]$.
For instance, given a fact $\atoma = p(\anull_x, \anull_y, \anull_x, const_1)$, $\canonical(\atoma)$ corresponds to the fact $\atoma^c = p(\anull^c_1, \anull^c_2, \anull^c_1, const_1)$.
Thus, the following results hold.

\begin{lemma}
\label{lemma:canonical-is-isomorphic}
Let $\atoma$ be a fact and $\atoma^c$ its $\canonical(\atoma)$.
Then there exists an isomorphism between $\atoma$ and $\atoma^c$.
\end{lemma}

\begin{proposition}
    \label{prop:isomorphic-streaming}
    Let $\atoma$, $\atomb$ be two facts and $\atoma^c$, $\atomb^c$ their canonical form, respectively.
    Then $\atoma$ is isomorphic to $\atomb$ iff $\atoma^c = \atomb^c$.
    \end{proposition}

\noindent \textbf{Algorithm Overview.}
~\Cref{def:null-canonicalization} and~\Cref{prop:isomorphic-streaming} provide a powerful foundation to reduce the isomorphism-based firing condition to an equivalence check between hashes of facts by means of canonicalization.
A simple and yet effective \textit{hash-based} firing condition $\isomorphismcheckstreaming$ is provided in~\Cref{alg:isomorphismcheckstreaming-firing-condition}.

\begin{wrapfigure}[7]{L}{0.6\textwidth}
\vspace{-0.8cm}
{\begin{minipage}{\linewidth}
\begin{algorithm}[H]
    	\begin{algorithmic}[1]
    	    \scriptsize
    \Function{HashIso\_FiringCondition}{$\atoma$} 
    	    \Let{$\atoma^c$}{$\canonical(\atoma)$}
            \Let{$\hashset$}{$\mathsf{getGeneratedFactsHashSet}$()}
            \Let{\textit{hash}$_{\atoma^c}$}{$\mathsf{createStrongHash}$($\atoma^c$)}
    	    \If{$\hashset.\mathsf{contains}(\textit{hash}_{\atoma^c})$}\label{alg:isomorphismcheckstreaming:hashset-contains} \Return \textbf{false}
            \EndIf
            \State $\hashset.\mathsf{add}(\textit{hash}_{\atoma^c})$
    	    \State \Return \textbf{true}
    	\EndFunction
    \end{algorithmic}
\caption{$\isomorphismcheckstreaming$.}
\label{alg:isomorphismcheckstreaming-firing-condition}
\end{algorithm}
\end{minipage}}
\end{wrapfigure}

\noindent
The procedure employs a $\hashset$, i.e., a set data structure storing hash values, to check for membership. 
Note that there might be false positives, i.e. facts that should have been generated, but they are not generated because the check mistakenly detects them as isomorphic to already generated ones. However, using a \textit{hash function with strong collision resistance} should make this event improbable, as witnessed in practice.
Whenever an applicable homomorphism for a fact $\atoma$ is considered, it first computes $\canonical(\atoma)=\atoma^c$, and then checks whether the $\hashset$ contains such a newly generated fact (line~\ref{alg:isomorphismcheckstreaming:hashset-contains}).
If that is the case, then $\atoma^c$ has already been added in a previous iteration by another fired homomorphism over a fact $\atomb$, s.t. $\canonical(\atoma)=\atoma^c=\canonical(\atomb)$.
By~\Cref{prop:isomorphic-streaming}, this entails that $\atoma$ and $\atomb$ are isomorphic, thus the isomorphism check returns \textit{false} (as the firing condition is not satisfied).
Instead, if $\atoma^c$ is not contained in $\hashset$, then it updates $\hashset$ with the new hash and returns \textit{true}.

Note that the amortized running time complexity of the hash-based isomorphism check is $\O(1)$ if the set is implemented using hash tables~\cite{clrs-book}, which makes such a firing condition particularly suitable for big data and streaming settings.
If the occurrence of collisions (i.e. two facts with the same hash value) could not be completely avoided, a collision resolution must be performed.
Nevertheless, if a strong hash function is employed, this occurs very rarely in practice.
With reference to Example~\ref{ex:running-example}, the canonicalization of \textit{WorksFor}(\textit{Bob},$\nu_1$) is \textit{WorksFor}(\textit{Bob},$\phi_1$), which is the same as \textit{WorksFor}(\textit{Bob},$\nu_2$) already in the chase, and the generation of the former is prevented by Algorithm~\ref{alg:isomorphismcheckstreaming-firing-condition}.
\section{The Chase in Streaming-based Architectures}
\label{sec:section4}
We now discuss the integration of the chase in streaming-based architectures.
We recall that in such systems facts can only be materialized at the end of the reasoning evaluation and, ideally, only the chase steps that are required to answer the query will be activated, without generating the full output of the procedure.
While we consider Vadalog as reference, the approach is generic for reasoners based on similar architectural principles.

\smallskip \noindent \textbf{Vadalog Streaming Architecture.}
First, we briefly illustrate how Vadalog streaming architecture operates.
The logic core of the system adheres to the \textit{pipes and filters}~\cite{buschmann2007pattern} data processing pattern and consists in an active pipeline that reads data from the input sources, performs the needed transformations, and produces the desired output as a result~\cite{BellomariniSG18}.
Given a database $D$, a set $\Sigma$ of Datalog$^\pm$ rules, and a query $q$, the system compiles a processing pipeline by adding a \textit{data scan} (a \textit{filter}) for each rule in $\Sigma$ and an edge (a \textit{pipe}) connecting a scan $\beta$ to a scan $\alpha$ if the head of rule $\beta$ unifies with a body atom of rule $\alpha$.
The reasoning is then performed in a pull-based query-driven fashion that generalizes the \textit{volcano iterator model}~\cite{GraefeM93}, where each scan (hence, each rule) reads facts from the respective parents, from the output scan corresponding to $q$ down to the data sources that inject ground facts from $D$ into the pipeline.
Interactions between scans occur via primitives such as $\mathsf{next()}$, which asks the parent scans whether there exists a fact to read, and employ specific \textit{routing strategies} to determine which interaction occurs first~\cite{BellomariniBGS22}.

\smallskip \noindent \textbf{Streaming Chase with Resumption.}
With the goal of enabling generic BCQA in a streaming-based architecture, we recall from Section~\ref{sec:section2} that resumption is required.
In fact, an alternative approach would consist in splitting the conjunctive query into an atomic one and a rule, which is then included in the input set $\Sigma$ of rules.
However, it is often unfeasible in practice, as will be demonstrated in Section~\ref{sec:section5}, due to the restrictions that would be required to limit the new rule to the syntax of $\Sigma$'s fragment.
For instance, if the query were to feature an attacked harmful join, the AHJE rewriting would be required, which however causes the generation of a number of Protected rules that is in the worst case exponential to the number of attacked harmful joins.

\begin{wrapfigure}[13]{L}{0.6\textwidth}
\vspace{-1.6cm}
{\begin{minipage}{\linewidth}
\begin{algorithm}[H]
    \scriptsize
    \begin{algorithmic}[1]
        \Function{chase$_s$}{$\dataset,\Sigma,\query$}
            \Let{$\mathit{maxRes}$}{$|\vars(\query)| + 1$}
            \Let{$p$}{$\mathsf{compile\text{-}pipeline}(\dataset,\Sigma,\query)$} \label{line:chase:pipeline}
            \While{$\mathsf{next}(p)$}
                \Let{$\langle s,f \rangle$}{$\mathsf{get}(p)$} \label{line:chase:applhom_start}
                \If{$\mathsf{isApplicableHomomorphism}(\langle s,f \rangle)$}
                    \Let{$\theta$}{$\mathsf{applicableHomomorphism}(\langle s,f \rangle)$} \label{line:chase:applhom_end}
                    \Let{$\atoma$}{$\fire(s,\theta)$} \label{line:chase:fire} 
                    \If{$\firingcondition(\atoma)$} \label{line:chase:fire-condition}
            	        \State $s.\mathsf{addFact}(\atoma)$
                        \If{$\mathsf{answer}(q)$} \Return \textbf{true} \label{line:chase:query}
                        \EndIf
                    \EndIf
                    \If{\textbf{not} $\mathsf{frozen}(\atoma)$ \textbf{and} $\atoma.\mathit{resIt} < \mathit{maxRes}$} \label{line:chase:res}
                        \Let{$\atoma$}{$\mathsf{freeze}(\atoma)$}
                        \Let{$\atoma.\mathit{resIt}$}{$\atoma.\mathit{resIt} + 1$} \label{line:chase:freeze}
                        \State \textbf{go to}~\Cref{line:chase:fire-condition} 
                    \EndIf
                \EndIf
            \EndWhile
            \State \Return \textbf{false}
        \EndFunction
	\end{algorithmic}    
\caption{The $\chases$ for streaming BCQA.}
\label{alg:streaming-chase}
\end{algorithm}
\end{minipage}}
\vspace{-1cm}
\end{wrapfigure}

\noindent
Standard resumption consists in performing full iterations of the chase and materializing the intermediate results, both prohibitive operations in a streaming environment.
Thus, we enable the Vadalog system to perform generic BCQA by developing a novel resumption-based chase procedure ($\chases$) for streaming environments, provided in Algorithm~\ref{alg:streaming-chase}.
Leveraging the pull-based processing pipeline approach, resumption is here treated as a \textit{fact-level property}, that is, each generated fact in the chase belongs to a specific iteration of resumption corresponding to the maximum between the iterations of its parent facts.
Thus, it can experience \textit{labelled null freezing}, up to the maximum number of resumptions $|\vars(q)| + 1$ allowed by the query $q$, without performing a global chase iteration.
Note that this approach is only applicable if $\Sigma$ does not feature attacked harmful joins, that is, in the context of Vadalog it belongs to Protected Datalog$^\pm$.
More specifically, given a database $D$, a set $\Sigma$ of Protected rules and a BCQ $\query$ which enables $\mathit{maxRes}$ iterations of resumption, first the procedure compiles the corresponding processing pipeline $p$ as discussed above, and stores it in a map of scans and parent scans (line~\ref{line:chase:pipeline}).
Then, at each iteration, it propagates via $\mathsf{next}$ the request to read facts from the query scan to its parents, down to the scans that read ground data from $D$, following a fixed routing strategy.
If the primitive returns \textit{true}, then it entails that there exists a scan $s$ in $p$ such that it can read a fact $f$ from its parents.
Thus, $f$ is read via $\mathsf{get}$ primitive and the algorithm checks whether an applicable homomorphism $\theta$ may occur from the rule corresponding to $\mathit{s}$ to $f$ (lines~\ref{line:chase:applhom_start}-\ref{line:chase:applhom_end}).
If that is the case, $\atoma$ is the fact derived from firing $\theta$ over $s$.
To determine whether $\atoma$ is generated in the chase, the procedure will make use of ad-hoc firing conditions (such as the streaming-friendly $\homomorphismcheckstreaming$ and $\isomorphismcheckstreaming$ discussed in Section~\ref{sec:section3}).
If the firing condition is satisfied, $\atoma$ is added to the set of facts that can be read from $s$.
Moreover, it checks whether the facts in $q$ are now able to provide a positive answer for the corresponding query, in which case the procedure terminates (line~\ref{line:chase:query}).
If the firing condition prevents the generation of $\atoma$ instead, then the fact-level resumption occurs.
Let $\mathit{resIt}$ be $\atoma$'s initial resumption iterations, derived from its parent facts.
If there exists at least a null in $\atoma$ that has not already been subjected to freezing and $\mathit{resIt} < \mathit{maxRes}$, then $\atoma$ is frozen, the frozen nulls are tagged as constants, and \textit{resIt} is increased (lines~\ref{line:chase:res}-\ref{line:chase:freeze}).
Finally, the firing condition check is performed again for the frozen $\atoma$.
If the facts from all the scans in $p$ have been read but a positive answer for $q$ was not achieved, then the procedure returns \textit{false}.

From an architectural perspective, this novel streaming-based chase with resumption is encapsulated in a \textit{termination wrapper} connected to each scan $s$.
It is responsible for enforcing the firing condition for each fact $\atoma$ resulting from an applicable homomorphism on $s$, as well as for performing null freezing via a \textit{resumption freezer} component.
Additionally, a \textit{query processor} is connected to the query scan and checks after each chase step whether $q$ has a positive answer.

\begin{figure*}[!t]
    \centering
    \begin{minipage}{0.45\textwidth}
    \centering
    \includegraphics[width=1.0\linewidth]{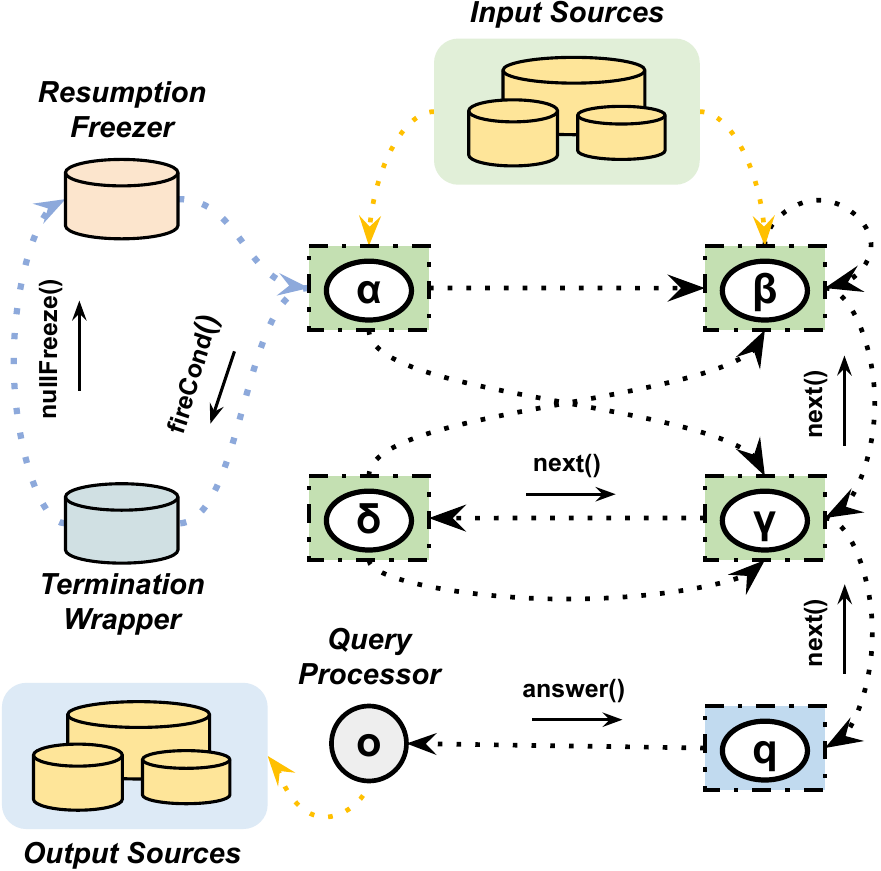}
    \end{minipage}
    \begin{minipage}{0.45\textwidth}
    \centering
    \includegraphics[width=1.2\linewidth]{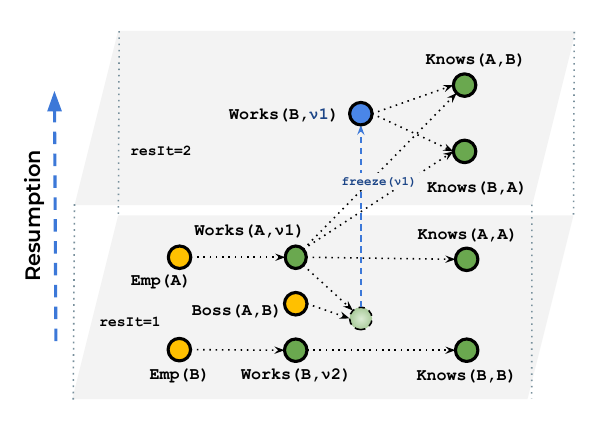}
    \end{minipage}
    \hspace{0.7cm}
    \caption{Vadalog processing pipeline and streaming chase for Example~\ref{ex:running-example}.}
    \label{fig:chases}
\vspace*{-5mm}
\end{figure*}
    
\smallskip \noindent \textbf{Streaming Chase over Example~\ref{ex:running-example}.}
To support the explanation of Algorithm~\ref{alg:streaming-chase}, we illustrate its application over Example~\ref{ex:running-example}, guided by a visual representation of Vadalog processing pipeline and the streaming chase resulting from a specific rule activation order in Figure~\ref{fig:chases}.
Note that, for space reasons, we do not explicitly address the rewriting of the attacked harmful join in rule $\gamma$ via AHJE to make $\Sigma$ Protected and enable the $\chases$ procedure.
Let us consider the BCQ $q: Q \leftarrow \textit{Knows}(\textit{Alice},x),\textit{Knows}(\textit{Bob},x)$, which asks whether there exists an $x$ who knows both \textit{Alice} and \textit{Bob} and enables up to $2$ iterations of resumption.
First, the processing pipeline is compiled, featuring a scan for $q$ and each rule in $\Sigma$, connected by the logical dependencies between them.
Then, the actual chase procedure begins with $\mathsf{next}$ calls from the scan $q$ to its parents, down to scan $\alpha$, which reads the facts \textit{Employee}(\textit{Alice}) and \textit{Employee}(\textit{Bob}) from the input sources and fires applicable homomorphisms to generate \textit{WorksFor}(\textit{Alice},$\nu_1$) and \textit{WorksFor}(\textit{Bob},$\nu_2$), respectively.
The new facts are in turn read by scan $\beta$, together with \textit{HasBoss}(\textit{Alice},\textit{Bob}) from the data sources.
The same scan attempts the generation of \textit{WorksFor}(\textit{Bob},$\nu_1$), which is however prevented by the termination wrapper as it is isomorphic with \textit{WorksFor}(\textit{Bob},$\nu_2$).
Thus, the freezing of \textit{WorksFor}(\textit{Bob},$\nu_1$) is performed by the dedicated component and $\nu_1$ is tagged as a constant.
Then, scan $\gamma$ reads the above facts and it generates \textit{Knows}(\textit{Alice},\textit{Alice}),\textit{Knows}(\textit{Bob},\textit{Bob}), \textit{Knows}(\textit{Alice},\textit{Bob}), and \textit{Knows}(\textit{Bob},\textit{Alice}).
Finally, $q$ reads the resulting facts and it returns a positive answer to the query processor, thus terminating the procedure.

\smallskip \noindent \textbf{Correctness and Firing Conditions.} 
As firing conditions for Algorithm~\ref{alg:streaming-chase} we employ the $\homomorphismcheckstreaming$ and $\isomorphismcheckstreaming$ presented in Section~\ref{sec:section3}.
Furthermore, this novel approach allowed us to integrate Shy into the Vadalog engine, thus supporting for the first time, to the best of our knowledge, both these powerful and expressive fragments in a streaming-based reasoning system.
We will delve into it in a future work.
We now argue the correctness of \Cref{alg:streaming-chase}.
\begin{theorem}
\label{thm:pchase-streaming}
    For any database $D$, a Shy set $\Sigma$ of rules, a BCQ $\query = \psi(\bar{z})$, we have that $\pchase_r(D,\Sigma) \models q$ iff $\chases(D,\Sigma)\models q$ with firing condition $\homomorphismcheckstreaming$.
\end{theorem}

\begin{theorem}
\label{thm:ichase-streaming}
    For any database $D$, a Shy set $\Sigma$ of rules, a BCQ $\query = \psi(\bar{z})$, we have that $\ichaser(D,\Sigma) \models q$ iff $\chases(D,\Sigma) \models q$ with firing condition $\isomorphismcheckstreaming$.
\end{theorem}
\section{Experimental Evaluation}
\label{sec:section5}
We experimentally compared the chase variants presented above over well-known scenarios with existential quantification in the context of benchmarking $\datalogpm$ reasoners.
The experiments were run on a cloud-based virtual machine equipped with a CPU Intel Xeon Platinum 8171M @ 2.60GHz (4 cores) and with $16$GB of RAM.
The results of the experiments, as well as the steps to reproduce them, were made available as supplementary material~\cite{material}, whereas the Vadalog system will be made available upon request.
Note that $\vadalogI$ is the Vadalog configuration based on $\isomorphismcheckstreaming$, $\vadalogP$ is the one based on $\homomorphismcheckstreaming$, $\vadalogIR$ and $\vadalogPR$ are the ones in streaming resumption mode for generic conjunctive query answering based on $\isomorphismcheckstreaming$ and $\homomorphismcheckstreaming$, respectively.
Thus, in the presence of a CQ, both $\vadalogI$ and $\vadalogP$ require splitting it into an atomic one and a rule, which is then integrated in the set $\Sigma$, possibly performing rewriting steps such as AHJE to restrict its syntax to the one of $\Sigma$.

\begin{figure}[t!]
\begin{center}
    \begin{minipage}{0.41\textwidth}
    \centering
    \includegraphics[width=1.0\linewidth]{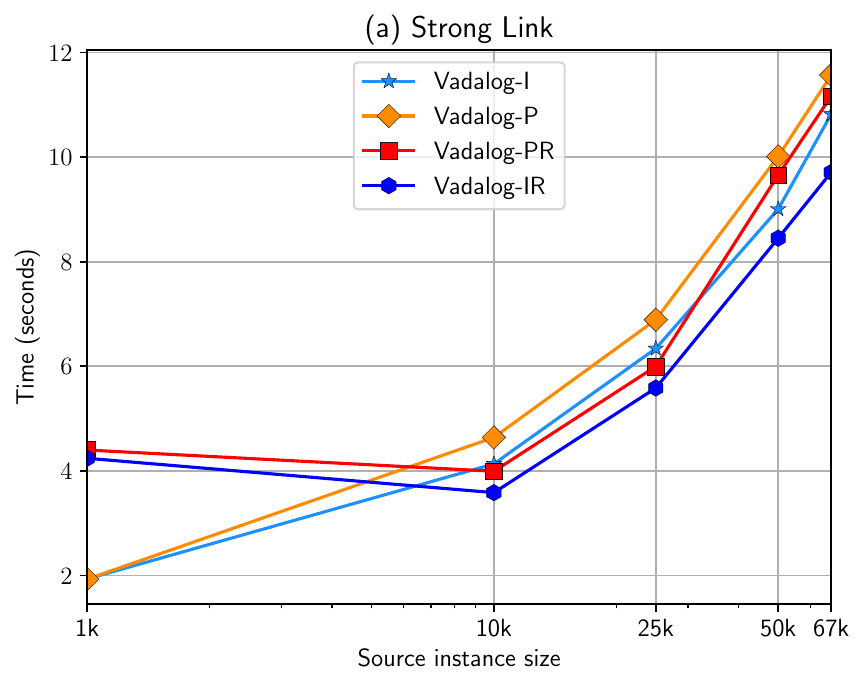}
    \end{minipage}
    \begin{minipage}{0.41\textwidth}
    \centering
    \includegraphics[width=1.0\linewidth]{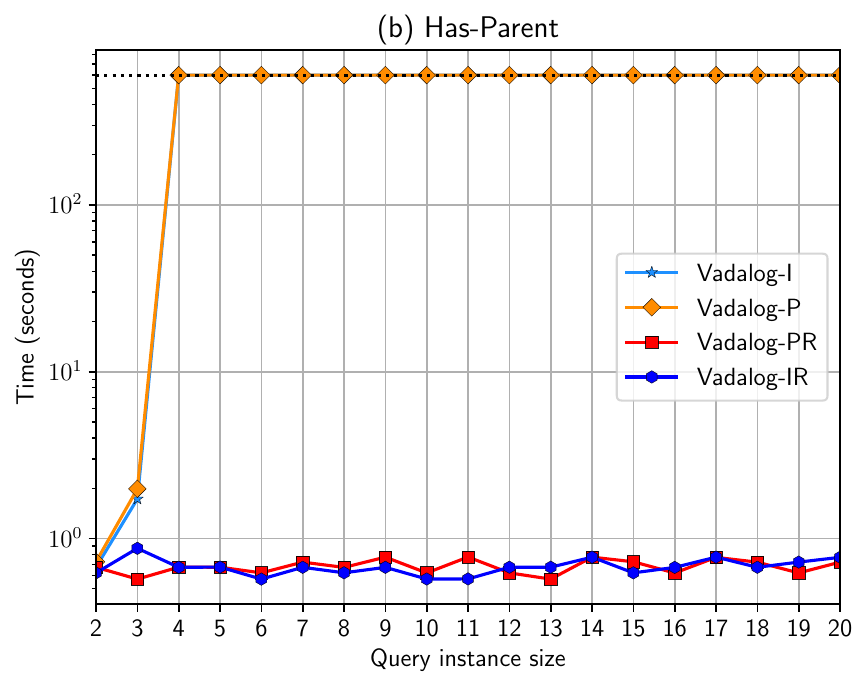}
    \end{minipage}
    \begin{minipage}{0.41\textwidth}
    \centering
    \includegraphics[width=1.0\linewidth]{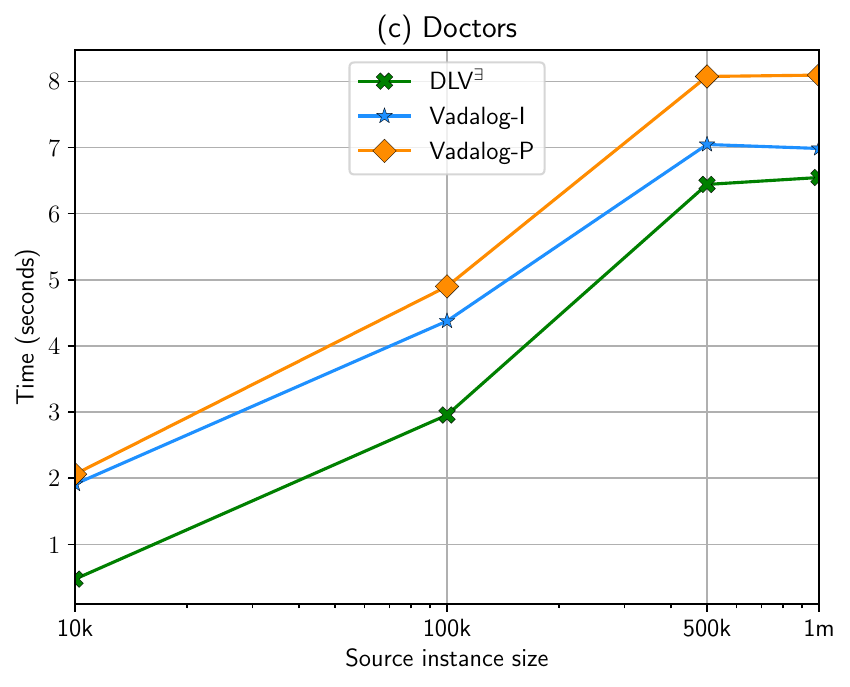}
    \end{minipage}
    \begin{minipage}{0.41\textwidth}
    \centering
    \includegraphics[width=1.0\linewidth]{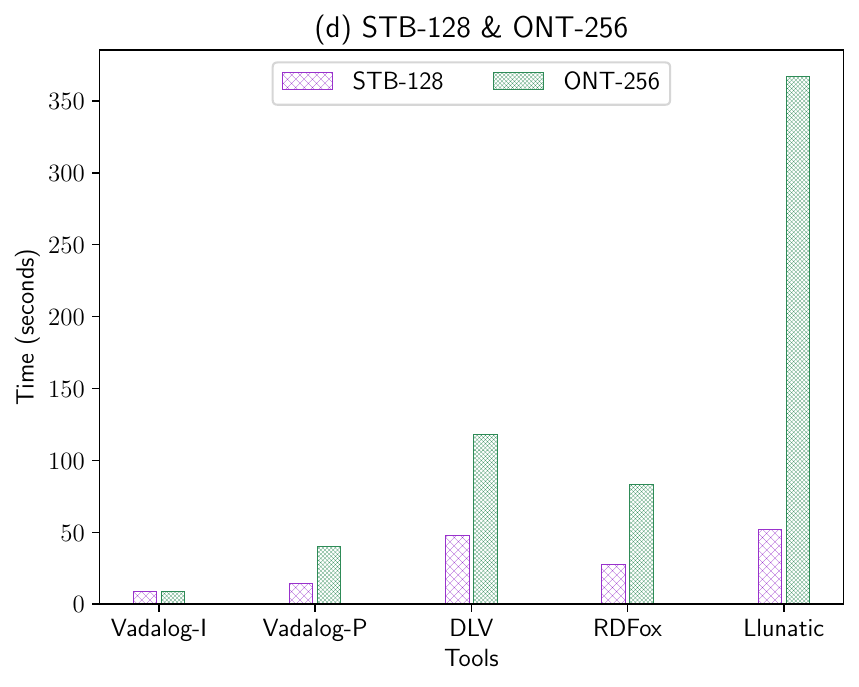}
    \end{minipage}
\end{center}
\caption{Reasoning times for the experimental evaluation.}
\label{fig:experiments} 
\end{figure}

\smallskip \noindent
\textbf{(a) Strong Link.}
The scenario is a variant of a financial recursive use case about relationships between companies~\cite[Example 3]{BellomariniSG18} over real data extracted from DBPedia~\cite{dbpedia}.
It consists in finding strong links between ``significantly controlled companies'', that is, companies for which there exist common significant shareholders (persons who hold more than $20\%$ of the stocks).
Together with~\ref{fig:experiments}(b), the goal of this experiment is to compare the performance of the four Vadalog configurations.
In this scenario, the query contains a simple attacked harmful join between two variables, which was included as a rule in the program via AHJE in the $\vadalogI$ and $\vadalogP$ configurations. 
We used source instances of $1K$, $10K$, $25K$, $50K$ and $67K$ companies and we ran the experiment for $10$ iterations, averaging the elapsed times.
As shown in Figure~\ref{fig:experiments}(a), the configurations achieve comparable performances.
The overhead introduced by the aggregate-fact tree in $\vadalogP$ and $\vadalogPR$ does not significantly influence this setting, being balanced out by the higher number of intermediate facts generated in $\vadalogI$ and $\vadalogIR$, and neither does the AHJE in $\vadalogI$ and $\vadalogP$, due to the simplicity of the attacked harmful join.

\smallskip \noindent
\textbf{(b) Has-Parent.}
The scenario is the simple set of rules: $r_1: Person(x)\to \exists y \, HasParent(x,y)$ and $r_2: HasParent(x,y)\to person(y)$.
It consists in running the following query $q: Q_n(x) \leftarrow person(x_0), hasParent(x_0,x_1), hasParent\\(x_1,x_2),$ $\dots, hasParent(x_{n-2},x_{n-1})$, where $n$ is the scaling parameters, over a database with a single fact: $D=\{person(Alice)\}$.
Note that, similarly to~\ref{fig:experiments}(a), the query is conjunctive and it features an attacked harmful join, therefore both $\vadalogI$ and $\vadalogP$ require to perform the AHJE rewriting.
However, as shown in Figure~\ref{fig:experiments}(b), $\vadalogI$ and $\vadalogP$ do not scale, timing out from $n=4$ onwards due to the AHJE step, which could not handle the increasing number of attacked harmful join to rewrite~\cite{BaldazziBSA21,BaldazziBSA22}.
On the other hand, both $\vadalogIR$ and $\vadalogPR$ achieve very low running times, almost constant for all values of the scaling parameter.
This experiment highlights the importance of integrating the resumption in Vadalog to keep the query unaffected by the limits of the fragment, as well as the general effectiveness of the implemented resumption approaches when the query rewriting becomes unfeasible.

\smallskip \noindent
\textbf{(c) Doctors.}
The scenario is a data integration task from the schema mapping literature~\cite{MeccaPS14}.
While non-recursive, it is rather important as a plausible real-world case.
In this experiment, we compare the performance of $\vadalogI$, $\vadalogP$ and the materialization-based $\dlve$~\cite{dlvE}.
Note that the resumption-based configurations are not included since the maximum number of resumption iterations for the queries is $1$.
We used source instances of $10K$, $100K$, $500K$, $1M$ facts and we ran $9$ queries for $10$ iterations, averaging the elapsed times.
As shown in Figure~\ref{fig:experiments}(c), $\dlve$ performs slightly better than the two configurations of Vadalog over smaller instances.
On the other hand, the gap considerably shrinks with the increase in the dataset size. 
This can be explained by the time spent by Vadalog to perform program optimizations before the actual execution of the reasoning.
Regarding $\vadalogI$ and $\vadalogP$, we observe that once again there is no heavy performance degradation due to the adoption of $\homomorphismcheckstreaming$.

\smallskip \noindent
\textbf{(d) STB-128 \& ONT-256.}
The scenarios are featured in the \textit{ChaseBench}\cite{BenediktKMMPST17} benchmarks.
Specifically, \textit{STB-128} is a set of about $250$ Warded rules, $25\%$ of which contain existentials.
The expected target instance contains $800K$ facts, with $20$ of labelled nulls.
We ran $16$ different queries for $10$ iterations, averaging the elapsed times.
Queries are rather complex, as they involve on average $5$ joins, and $8$ of them are attacked harmful.
On the other hand, \textit{ONT-256} is a set of $789$ Warded rules, $35\%$ of which contain existentials.
Rules are even more complex than STB-128, and contain multiple joins as well as pervasive recursion.
The expected target instance contains $~2$ million facts, with an incidence of $50\%$ of labelled nulls.
We ran $10$ different queries, averaging the elapsed times.
Queries involve an average of $5$ joins, and $5$ of them are attacked harmful.
In this experiment, we compare the performance of $\vadalogI$, $\vadalogP$, DLV~\cite{LeonePFEGPS06}, $\rdfox$~\cite{NenovPMHWB15}, and $\llunatic$~\cite{GeertsMPS14}.
Note that $\vadalogIR$ and $\vadalogPR$ were not tested as the queries did not require resumption, and that AHJE is performed, on the queries with attacked harmful joins, without significantly affecting the performance.
As shown in Figure~\ref{fig:experiments}(d), Vadalog, both in $\vadalogI$ and $\vadalogP$, outperforms all the other systems in both scenarios: $\vadalogI$ and $\vadalogP$ ran in $8.88$ and $14.34$ seconds for \textit{STB-128} and in $8.86$ and $40.24$ seconds for \textit{ONT-256}.
Indeed, $\vadalogI$ is on average $3$ times faster than $\rdfox$, and $7$ times faster than $\llunatic$.
\section{Conclusion}
\label{sec:conclusion}
To perform QA tasks over large and complex datasets, modern $\datalogpm$ reasoners such as Vadalog shift from a materialization approach to a streaming-based one, which enables scalable query-driven evaluations.
In this paper, we focus on the expressive Shy and Warded fragments of $\datalogpm$.
Leveraging their theoretical underpinnings, we develop variants of the chase procedure that are highly suited for streaming-based architectures, and we integrate them in Vadalog to efficiently answer generic conjunctive queries in complex settings.
\begin{credits}
\subsubsection{\ackname}
This work was partially supported by the Vienna Science and Technology Fund (WWTF) [10.47379/ICT2201, 10.47379/VRG18013, 10.47379/\\NXT22018]; and the Christian Doppler Research Association (CDG) JRC LIVE.

\subsubsection{\discintname}
The authors have no competing interests to declare that are
relevant to the content of this article.

\end{credits}
%
%
%
%
\bibliographystyle{splncs04}
\bibliography{biblio}

\begin{thebibliography}{10}
\providecommand{\url}[1]{\texttt{#1}}
\providecommand{\urlprefix}{URL }
\providecommand{\doi}[1]{https://doi.org/#1}

\bibitem{AbiteboulHV95}
Abiteboul, S., Hull, R., Vianu, V.: Foundations of Databases (1995)

\bibitem{DBLP:conf/jelia/AmendolaM19}
Amendola, G., Marte, C.: Extending bell numbers for parsimonious chase estimation. In: {JELIA}. Lecture Notes in Computer Science, vol. 11468, pp. 490--497. Springer (2019)

\bibitem{material}
Baldazzi, T., Bellomarini, L., Favorito, M., Sallinger, E.: Supplementary material. \url{https://bit.ly/47DHCTS}, accessed: 2023-11-20

\bibitem{BaldazziBFS22}
Baldazzi, T., Bellomarini, L., Favorito, M., Sallinger, E.: On the relationship between shy and warded datalog+/-. In: {KR} (2022)

\bibitem{BaldazziBSA21}
Baldazzi, T., Bellomarini, L., Sallinger, E., Atzeni, P.: Eliminating harmful joins in warded datalog+/-. In: RuleML+RR. Lecture Notes in Computer Science, vol. 12851, pp. 267--275. Springer (2021)

\bibitem{BaldazziBSA22}
Baldazzi, T., Bellomarini, L., Sallinger, E., Atzeni, P.: Reasoning in warded datalog+/- with harmful joins. In: {SEBD}. {CEUR} Workshop Proceedings, vol.~3194, pp. 292--299. CEUR-WS.org (2022)

\bibitem{BellomariniBBS22}
Bellomarini, L., Benedetto, D., Brandetti, M., Sallinger, E.: Exploiting the power of equality-generating dependencies in ontological reasoning. Proc. {VLDB} Endow.  \textbf{15}(13),  3976--3988 (2022)

\bibitem{BellomariniBGS22}
Bellomarini, L., Benedetto, D., Gottlob, G., Sallinger, E.: Vadalog: {A} modern architecture for automated reasoning with large knowledge graphs. IS  \textbf{105} (2022)

\bibitem{BellomariniGPS17}
Bellomarini, L., Gottlob, G., Pieris, A., Sallinger, E.: Swift logic for big data and knowledge graphs. In: {IJCAI}. pp. 2--10. ijcai.org (2017)

\bibitem{BellomariniSG18}
Bellomarini, L., Sallinger, E., Gottlob, G.: The vadalog system: Datalog-based reasoning for knowledge graphs. Proc. {VLDB} Endow.  \textbf{11}(9),  975--987 (2018)

\bibitem{BenediktKMMPST17}
Benedikt, M., Konstantinidis, G., Mecca, G., Motik, B., Papotti, P., Santoro, D., Tsamoura, E.: Benchmarking the chase. In: {PODS}. pp. 37--52. {ACM} (2017)

\bibitem{buschmann2007pattern}
Buschmann, F., Henney, K., Schmidt, D.C.: Pattern-Oriented Software Architecture, A Pattern Language for Distributed Computing, vol.~4 (2007)

\bibitem{CaliGK13}
Cal{\`{\i}}, A., Gottlob, G., Kifer, M.: Taming the infinite chase: Query answering under expressive relational constraints. J. Artif. Intell. Res.  \textbf{48},  115--174 (2013)

\bibitem{CaliGL12}
Cal{\`{\i}}, A., Gottlob, G., Lukasiewicz, T.: A general datalog-based framework for tractable query answering over ontologies. J. Web Semant.  \textbf{14},  57--83 (2012)

\bibitem{CaliGLMP10}
Cal{\`{\i}}, A., Gottlob, G., Lukasiewicz, T., Marnette, B., Pieris, A.: Datalog+/-: {A} family of logical knowledge representation and query languages for new applications. In: {LICS}. pp. 228--242. {IEEE} Computer Society (2010)

\bibitem{CaliGP10}
Cal{\`{\i}}, A., Gottlob, G., Pieris, A.: Advanced processing for ontological queries. Proc. {VLDB} Endow.  \textbf{3}(1),  554--565 (2010)

\bibitem{clrs-book}
Cormen, T.H., Leiserson, C.E., Rivest, R.L., Stein, C.: Introduction to Algorithms, 3rd Edition. {MIT} Press (2009)

\bibitem{dbpedia}
DBpedia: Web site. \url{https://www.dbpedia.org} (2018), accessed: 2023-11-20

\bibitem{FaginKMP05}
Fagin, R., Kolaitis, P.G., Miller, R.J., Popa, L.: Data exchange: semantics and query answering. Theor. Comput. Sci.  \textbf{336}(1),  89--124 (2005)

\bibitem{GeertsMPS14}
Geerts, F., Mecca, G., Papotti, P., Santoro, D.: That's all folks! {LLUNATIC} goes open source. Proc. {VLDB} Endow.  \textbf{7}(13),  1565--1568 (2014)

\bibitem{GottlobP15}
Gottlob, G., Pieris, A.: Beyond {SPARQL} under {OWL} 2 {QL} entailment regime: Rules to the rescue. In: {IJCAI}. pp. 2999--3007. {AAAI} Press (2015)

\bibitem{GraefeM93}
Graefe, G., McKenna, W.J.: The volcano optimizer generator: Extensibility and efficient search. In: {ICDE}. pp. 209--218. {IEEE} Computer Society (1993)

\bibitem{JohnsonK84}
Johnson, D.S., Klug, A.C.: Testing containment of conjunctive queries under functional and inclusion dependencies. J. Comput. Syst. Sci.  \textbf{28}(1),  167--189 (1984)

\bibitem{KrotzschT16}
Kr{\"{o}}tzsch, M., Thost, V.: Ontologies for knowledge graphs: Breaking the rules. In: {ISWC} {(1)}. Lecture Notes in Computer Science, vol.~9981, pp. 376--392 (2016)

\bibitem{dlvE}
Leone, N., Manna, M., Terracina, G., Veltri, P.: {DLV}\^{}{E} system. \url{https://www.mat.unical.it/dlve/} (2017), accessed: 2023-11-20

\bibitem{LeoneMTV19}
Leone, N., Manna, M., Terracina, G., Veltri, P.: Fast query answering over existential rules. {ACM TOCL}  \textbf{20}(2),  12:1--12:48 (2019)

\bibitem{LeonePFEGPS06}
Leone, N., Pfeifer, G., Faber, W., Eiter, T., Gottlob, G., et. al.: The {DLV} system for knowledge representation and reasoning. {ACM TOCL}  \textbf{7}(3),  499--562 (2006)

\bibitem{MaierMS79}
Maier, D., Mendelzon, A.O., Sagiv, Y.: Testing implications of data dependencies. {ACM TODS}  \textbf{4}(4),  455--469 (1979). \doi{10.1145/320107.320115}

\bibitem{MeccaPR12}
Mecca, G., Papotti, P., Raunich, S.: Core schema mappings: Scalable core computations in data exchange. Inf. Syst.  \textbf{37}(7),  677--711 (2012)

\bibitem{MeccaPS14}
Mecca, G., Papotti, P., Santoro, D.: {IQ-METER} - an evaluation tool for data-transformation systems. In: {ICDE}. pp. 1218--1221. {IEEE} Computer Society (2014)

\bibitem{NenovPMHWB15}
Nenov, Y., Piro, R., Motik, B., Horrocks, I., Wu, Z., Banerjee, J.: Rdfox: {A} highly-scalable {RDF} store. In: {ISWC} {(2)}. LNCS, vol.~9367, pp. 3--20. Springer (2015)

\bibitem{Pitoura2018}
Pitoura, E.: Pipelining, pp. 2768--2768. Springer New York, New York, NY (2018). \doi{10.1007/978-1-4614-8265-9\_872}

\end{thebibliography}
\newpage
\appendix
\newtheorem{mytheorem}{Theorem}
\numberwithin{mytheorem}{section}
\newtheorem{mylemma}{Lemma}
\numberwithin{mylemma}{section}
\newtheorem{mydefinition}{Definition}
\numberwithin{mydefinition}{section}
\newtheorem{myproposition}{Proposition}
\numberwithin{myproposition}{section}

\section{Chase-based Reasoning over Shy and Warded}

We collect existing results and present new theory regarding the applicability, termination, and correctness of the parsimonious and isomorphic chase variants over Shy, Warded, and their recently introduced intersection fragment Protected~\cite{BaldazziBFS22}, both for Boolean atomic (BAQA) and conjunctive (BCQA) query answering.
Figure~\ref{tab:summary} summarizes our results, presented later in this section.
At the essence of them, we have that both $\pchase$ and $\ichase$ are applicable to generic BAQA and BCQA tasks over Shy and, consequently, Protected.
On the other hand, neither $\pchase$ nor $\ichase$ can be applied to Warded, as this fragment allows a form of ``harmful'' join between nulls, which affects QA correctness.

\vspace{-0.5cm}

\begin{figure}[H]
    \small
    \centering
    {\renewcommand{\arraystretch}{2}%
    \begin{tabular}{|c|c|c|c|}
        \hline
         & Shy & Warded & Protected \\\hline
         \pchase & \makecell{\cmark (only atomic)\\Theorem~\ref{thm:baqa-Shy-pchase}} & \makecell{\xmark\\Theorem~\ref{thm:baqa-Warded-pchase}} & \makecell{\cmark (only atomic)\\Theorem~\ref{thm:baqa-Protected-pchase}} \\\hline
         \ichase & \makecell{\cmark (only atomic)\\Theorem~\ref{thm:baqa-Shy-ichase}} & \makecell{\xmark\\Theorem~\ref{thm:baqa-Warded-ichase}} & \makecell{\cmark (only atomic)\\Theorem~\ref{thm:baqa-Protected-ichase}} \\\hline
         \pchaser & \makecell{\cmark\\Theorem~\ref{thm:bcqa-Shy-pchaser}} & \makecell{\xmark\\Theorem~\ref{thm:bcqa-Warded-pchaser}} & \makecell{\cmark\\Theorem~\ref{thm:bcqa-Protected-pchaser}} \\\hline
         \ichaser & \makecell{\cmark\\Theorem~\ref{thm:bcqa-Shy-ichaser}} & \makecell{\xmark\\Theorem~\ref{thm:bcqa-Warded-ichaser}} & \makecell{\cmark\\Theorem~\ref{thm:bcqa-Protected-ichaser}} \\\hline
    \end{tabular}}
    \vspace{0.3cm}
    \caption{Results for $\pchase$, $\ichase$, $\pchaser$ and $\ichaser$, for BQA over Shy, Warded and Protected Datalog$^\pm$. The checkmark ``\cmark'' means that the chase procedure is sound and complete for Boolean QA over the fragment; otherwise we use a ``\xmark''.}
    \label{tab:summary}
\end{figure}

\noindent
Observe that for any database $D$, $\datalogpm$ set $\Sigma$ of rules and query $\query$, $\pchase(D,\Sigma)$ $\subseteq\ichase(D,\Sigma)$, since $\homomorphismcheck$ is a stricter firing condition than $\isomorphismcheck$, and that they are finite (~\cite[Prop. 3.5]{LeoneMTV19} and~\cite[Thm. 3.14]{BellomariniBBS22}, respectively).
From this, we report the following results~\cite{BaldazziBFS22}.

\begin{mylemma}
\label{lemma:pchase-subseteq-ichase}
    For any database $D$, $\datalogpm$ set $\Sigma$ of rules and query $\query$, $\pchase(D,\Sigma)$ $\subseteq\ichase(D,\Sigma)$.
\end{mylemma}
\begin{proof}
    Intuitively, we observe that Lemma~\ref{lemma:pchase-subseteq-ichase} holds because $\homomorphismcheck$ is a stricter firing condition than $\isomorphismcheck$, thus more applicable homomorphisms satisfy the $\ichase$ rather than the $\pchase$.    
\end{proof}

\noindent
We also consider the finiteness of the chase variants.

\begin{mylemma}\label{lemma:pchase-termination}
    Given a database $D$, a $\datalogpm$ set of rules and a query $\query$, $\pchase(D,\Sigma)$ is finite.
\end{mylemma}
\begin{proof}
    It follows from~\cite[Proposition 3.5]{LeoneMTV19}.
\end{proof}

\begin{mylemma}\label{lemma:ichase-termination}
    Given a database $D$, a $\datalogpm$ set of rules and a query $\query$, $\ichase(D,\Sigma)$ is finite.
\end{mylemma}
\begin{proof}
    The result follows as a corollary from~\cite[Theorem 3.14]{BellomariniBBS22}.
\end{proof}

\subsection{Boolean Atomic QA with the Chase}
Observe that both the parsimonious and the isomorphic variants of the chase proved to work only to answer atomic queries over Shy and Warded, respectively, whereas they do not ensure correctness for generic conjunctive queries~\cite{LeoneMTV19,BellomariniBGS22}.
We now investigate the applicability of $\pchase$ and $\ichase$ over the Shy and the Warded fragments for BAQA.

\smallskip \noindent \textbf{BAQA with Chase over Shy.}
Let us first consider Shy.
We recall the following result~\cite[Theorem 3.6]{LeoneMTV19}.

\begin{mytheorem}[BAQA with $\pchase$ over Shy]
\label{thm:baqa-Shy-pchase}
    Given a Shy set $\Sigma$ of rules, a database $D$, and a BA query $\query$, $\ochase(D,\Sigma)\models\query \text{ iff } \pchase(D,\Sigma)\models\query$.
\end{mytheorem}

\noindent
Regarding $\ichase$, we prove the following theorem.

\begin{mytheorem}[BAQA with $\ichase$ over Shy]
\label{thm:baqa-Shy-ichase}
    Given a set $\Sigma$ of Shy rules, a database $D$, and a BA query $\query$, $\ochase(D,\Sigma)\models\query \text{ iff } \ichase(D,\Sigma)\models\query$.
\end{mytheorem}
\begin{proof}
    Let the atomic query $q = \atoma$.
    Termination is guaranteed by Lemma~\ref{lemma:ichase-termination}. 
    The ``if'' direction (\textit{soundness}) holds because by construction $\ichase(D,\Sigma)\subseteq\ochase(D,\Sigma)$, for any database $D$ and program $\Sigma$. 
    Regarding the ``only if'' direction (\textit{completeness}), by assumption we know that there is a homomorphism $\theta$ from $\{\atoma\}$ to $\ochase(D,\Sigma)$.
    However, by definition of parsimonious programs~\cite[Definition 3.4]{LeoneMTV19}, we also know that there is a homomorphism $\theta'$ from $\{\theta(\atoma)\}$ to $\pchase(D,\Sigma)$.
    Hence, $\theta' \circ \theta$ is a homomorphism from $\{\atoma\}$ to $\pchase(D,\Sigma)$.
    By Lemma~\ref{lemma:pchase-subseteq-ichase}, since $\pchase(D,\Sigma)\subseteq\ichase(D,\Sigma)$, $\theta'$ is also a homomorphism from $\{\theta(\atoma)\}$ to $\ichase(D,\Sigma)$, thus proving the claim.
\end{proof}

\smallskip \noindent \textbf{BAQA with Chase over Protected.}
By definition of Protected as the intersection fragment between Shy and Warded~\cite[Theorem 1]{BaldazziBFS22}, the same results as Shy hold.

\begin{mytheorem}[BAQA with $\ichase$ over Protected]
\label{thm:baqa-Protected-ichase}
Given a Protected set $\Sigma$ of rules, a database $D$, and a boolean atomic query $\query$, we have that $\ochase(D,\Sigma)$ $\models\query \text{ iff } \ichase(D,\Sigma)\models\query$.
\end{mytheorem}

\begin{mytheorem}[BAQA with $\pchase$ over Protected]
\label{thm:baqa-Protected-pchase}
    Given a Protected set $\Sigma$ of rules, a database $D$, and a boolean atomic query $\query$, $\ochase(D,\Sigma)\models\query \text{ iff } \\\pchase(D,\Sigma)\models\query$.
\end{mytheorem}

\smallskip \noindent \textbf{BAQA with Chase over Warded.}
Let us now discuss BCQA over the Warded fragment.
To achieve this, we make use of~\Cref{ex:running-example} previously introduced.

\begin{mytheorem}[BAQA with $\ichase$ over Warded]
\label{thm:baqa-Warded-ichase}
There exist a database $D$, a Warded set $\Sigma$ of rules, and an atomic query $q$ such that BAQA with $\ichase$ is sound but not complete, i.e. $\ochase(D,\Sigma)\models q \nRightarrow \ichase(D,\Sigma)\models q$.
\end{mytheorem}
\begin{proof}
    Soundness follows from definition of $\ichase$, since $\isomorphismcheck$ is a stricter firing condition than $\ochase$. 
    We disprove completeness by counterexample via~\Cref{ex:running-example}.
    The result of computing $\ichase(D,\Sigma)$ is $D$ $\cup$ \{\textit{WorksFor}\\(\textit{Alice},$\nu_1$), \textit{WorksFor}(\textit{Bob},$\nu_2$), \textit{Knows}(\textit{Alice},\textit{Alice}), \textit{Knows}(\textit{Bob},\textit{Bob}\}.
    Note that the fact \textit{WorksFor}(\textit{Bob},$\nu_1$) is not generated in $\ichase(D,\Sigma)$, as there is a isomorphic embedding of it to \textit{WorksFor}(\textit{Bob},$\nu_2$).
    Let $\query=\textit{Knows}(\textit{Alice},\textit{Bob})$.
    Indeed, $\ichase(D,\Sigma)\not\models \query$, therefore the claim holds.
\end{proof}

\begin{mytheorem}[BAQA with $\pchase$ over Warded]
\label{thm:baqa-Warded-pchase}
There exist a database $D$, a Warded set $\Sigma$ of rules, and an atomic query $q$ such that BAQA with $\pchase$ is sound but not complete, i.e. $\ochase(D,\Sigma)\models q \nRightarrow \pchase(D,\Sigma)\models q$.
\end{mytheorem}
\begin{proof}
    Soundness follows from Lemma~\ref{lemma:pchase-subseteq-ichase}.
    We disprove completeness by counterexample via~\Cref{ex:running-example}.
    The result of computing $\pchase(D,\Sigma)$ is $D$ $\cup$ \{\textit{Works For}(\textit{Alice},$\nu_1$), \textit{WorksFor}(\textit{Bob},$\nu_2$), \textit{Knows}(\textit{Alice},\textit{Alice}), \textit{Knows}(\textit{Bob},\textit{Bob}\}.
    Note that the fact \textit{WorksFor}(\textit{Bob},$\nu_1$) is not generated in $\pchase(D,\Sigma)$, as there is a homomorphism from it to \textit{WorksFor}(\textit{Bob},$\nu_2$).
    Let $\query=\textit{Knows}(\textit{Bob},\textit{Alice})$.
   Indeed, $\pchase(D,\Sigma)\not\models \query$, therefore the claim holds.
\end{proof}

\subsection{Boolean Conjunctive QA with the Chase}
We recall the finiteness of the $\pchaser$.

\begin{mylemma}\label{lemma:pchaser-termination}
    Given a database $D$ and a $\datalogpm$ set of rules, $\pchaser(D,\Sigma, i)$ is finite for all $i\ge 0$.
\end{mylemma}
\begin{proof}
    It follows from~\cite[Theorem 4.9]{LeoneMTV19}.
\end{proof}

\noindent
We now investigate the applicability of $\pchase$ and $\ichase$ over the Shy and the Warded fragments for BCQA.

\smallskip \noindent \textbf{BCQA with Chase over Shy.}
Let us first consider Shy. We recall the result below~\cite[Theorem 4.11]{LeoneMTV19}.

\begin{mytheorem}[BCQA with $\pchaser$ over Shy]
\label{thm:bcqa-Shy-pchaser}
     Given a Shy set $\Sigma$, a database $D$, and a BCQ $\query$, $\ochase(D,\Sigma)\models\query \text{ iff } \pchaser(D,\Sigma,|\vars(q)| + 1)\models\query$.
\end{mytheorem}

\noindent
We can extend Theorem~\ref{thm:baqa-Shy-ichase} to perform BCQA with $\ichase$ over Shy, by introducing the \textit{isomorphic chase with resumption} ($\ichaser$), analogous to the $\pchaser$.
Formally, $\ichaser(D,\Sigma,0) = D$, and $\ichaser(D,\Sigma,i) = \ichase(\freeze{\ichaser$ $(D,\Sigma, i-1)}, \Sigma)$ for $i\ge 1$.
Considering this variant, we observe the following.

\begin{mylemma}\label{lemma:pchaser-subseteq-ichaser}
   For any database $D$, $\datalogpm$ set $\Sigma$ of rules, and $i\ge 0$, $\pchaser\\(D,\Sigma,i)\subseteq\ichaser(D,\Sigma,i)$.
\end{mylemma}
\begin{proof}
    It can be proved by induction on $i$.
    The base case holds by Lemma~\ref{lemma:pchase-subseteq-ichase}.
    The inductive case holds as $\pchaser(D,\Sigma, i) = \pchase(\freeze{\pchaser(D,\Sigma,i-1)}, \Sigma)$ by definition, $\pchase(\freeze{\pchaser(D,\Sigma,i-1)}, \Sigma)\subseteq \ichase(\freeze{\ichaser(D,\Sigma,i-1)},\Sigma)$ by monotonicity of the chase and by inductive hypothesis.
    The claim holds as $\ichase(\freeze{\ichaser(D,\Sigma,i-1)},\Sigma)$ is $\ichaser(D,\Sigma,i)$ by definition.
\end{proof}

\noindent
Note that the resumption introduces a structural dependency on the query in the chase.
Indeed, the maximum number of iterations for the resumption that can be performed to answer a certain BCQ $q$ depends on the query itself and it corresponds to the number of variables in the BCQ $|\vars(q)| + 1$.
Thus, we prove the following theorem.

\begin{mytheorem}[BCQA with $\ichaser$ over Shy]
\label{thm:bcqa-Shy-ichaser}
     Given a Shy set $\Sigma$, a database $D$, and a BCQ $\query$, $\ochase(D,\Sigma)\models\query \text{ iff } \ichaser(D,\Sigma,|\vars(q)| + 1)\models\query$.
\end{mytheorem}
\begin{proof}[Proof]
    Termination of $\ichaser$ is guaranteed by Lemma~\ref{lemma:ichase-termination} and by the limit in the number of iterations for the resumption.
    Soundness holds by construction as $\ichaser(D,\Sigma,|\vars(q)| + 1)\subseteq\ochase(D,\Sigma)$.
    Completeness directly follows from Theorem~\ref{thm:bcqa-Shy-pchaser} and Lemma~\ref{lemma:pchaser-subseteq-ichaser}.
\end{proof}

\smallskip \noindent \textbf{BCQA with Chase over Protected.}
By definition of Protected as the intersection fragment between Shy and Warded, the same results as Shy hold.

\begin{mytheorem}[BCQA with $\pchaser$ over Protected]
\label{thm:bcqa-Protected-pchaser}
     Given a Protected set $\Sigma$ of rules, a database $D$, and a Boolean conjunctive query $\query$, $\ochase(D,\Sigma)\models\query \text{ iff } \pchaser(D,\Sigma, |\vars(q)| + 1)\models\query$.
\end{mytheorem}

\begin{mytheorem}[BCQA with $\ichaser$ over Protected]
\label{thm:bcqa-Protected-ichaser}
     Given a Protected set $\Sigma$ of rules, a database $D$, and a Boolean conjunctive query $\query$, $\ochase(D,\Sigma)\models\query \text{ iff } \ichaser(D,\Sigma,|\vars(q)| + 1)\models\query$.
\end{mytheorem}

\smallskip \noindent \textbf{BCQA with Chase over Warded.}
Let us now discuss BCQA over the Warded fragment.
To achieve this, we make use of~\Cref{ex:running-example} previously introduced.

\begin{mytheorem}[BCQA with $\ichaser$ over Warded]
\label{thm:bcqa-Warded-ichaser}
    There exist a database $D$, a Warded set $\Sigma$ of rules, and a conjunctive query $q$ such that BCQA with $\ichaser$ is sound but not complete, i.e. $\ochase(D,\Sigma)\models q \nRightarrow \ichaser(D,\Sigma)\models q$.
\end{mytheorem}
\begin{proof}
    Soundness follows from the definition of resumption and of $\ichase$, since $\isomorphismcheck$ is a stricter firing condition than $\ochase$. 
    We disprove completeness by counterexample via~\Cref{ex:running-example}.
    Let $\query=\textit{Knows}(\textit{Alice},\textit{Bob}),\allowbreak \textit{Knows}(\textit{Bob},\textit{Alice})$.
    By definition of resumption, the maximum number of iterations is $1$.
    The result of computing $\ichaser(D,\Sigma,1)$ is $D$ $\cup$ \{\textit{WorksFor}(\textit{Alice},$\nu_1$), \textit{WorksFor}(\textit{Bob},$\nu_2$), \textit{Knows}(\textit{Alice},\textit{Alice}), \textit{Knows}(\textit{Bob},\textit{Bob}\}.
    Indeed, $\ichaser(D,\Sigma,\allowbreak 1)\allowbreak\not\models \query$, thus the claim holds.
\end{proof}

\begin{mytheorem}[BCQA with $\pchaser$ over Warded]
\label{thm:bcqa-Warded-pchaser}
    There exist a database $D$, a Warded set $\Sigma$ of rules, and a conjunctive query $q$ such that BCQA with $\pchaser$ is sound but not complete, i.e. $\ochase(D,\Sigma)\models q \nRightarrow \pchaser(D,\Sigma)\models q$.
\end{mytheorem}
\begin{proof}
    Soundness follows from Lemma~\ref{lemma:pchaser-subseteq-ichaser}. 
    We disprove completeness by counterexample via~\Cref{ex:running-example}.
    Let $\query=\textit{Knows}(\textit{Alice},\textit{Bob}), \textit{Knows}(\textit{Bob},\textit{Alice})$.
    By definition of resumption, the maximum number of iterations is $1$.
    The result of computing $\pchaser(D,\Sigma,1)$ is $D$ $\cup$ \{\textit{WorksFor}(\textit{Alice},$\nu_1$), \textit{WorksFor}(\textit{Bob},$\nu_2$), \textit{Knows}(\textit{Alice},\textit{Alice}), \textit{Knows}(\textit{Bob},\textit{Bob}\}.
    Indeed, $\pchaser(D,\Sigma,1)\not\models \query$, thus the claim holds.
\end{proof}

\setcounter{lemma}{0}
\setcounter{definition}{0}
\setcounter{proposition}{0}
\setcounter{theorem}{0}

\section{Streaming-friendly Firing Conditions in the Chase}

\begin{definition}
Given a predicate $p$ and a set $I$ of facts, an \emph{aggregate fact tree (af-tree)} $T_p$ is a tree s.t.: (i)~the root $T$ of $T_p$ is labelled by $[~]$; (ii)~a fact $\atoma = p(\bar{a}) = p(a_1,\dots,a_k) \in I$ iff there exists a path of $k$ nodes $n_{\bar{a}_1},\dots,n_{\bar{a}_k}$, where $n_{\bar{a}_i}$ ($1 \leq i \leq k$) is labelled by $[a_1,\dots,a_i]$, in $T_p$; and, (iii)~two nodes $n_{\bar{a}_j}$ and $n_{\bar{a}_{j+1}}$ $\in T_p$, labelled by $[a_1,\dots,a_j]$ and $[a_1,\dots,a_{j+1}]$ ($1\le j < k$), respectively, iff there exists an edge from $n_{\bar{a}_j}$ to $n_{\bar{a}_{j+1}}$ labelled by $a_{j+1}$.
\end{definition}

\begin{proposition}
    Let $\atoma = p(a_1,\dots,a_k)$ be a fact, $I$ a set of facts s.t. $\atoma \notin I$, and $T_p$ the af-tree for $\pred(\atoma) = p$ in $I$. Then there exists a homomorphism $\theta$ from $\atoma$ to $\atomb \in I$ iff there exists a root-to-leaf path $t$ in $T_p$ s.t. $\theta(\atoma[i]) = t[i]$, $1 \leq i \leq k$.
\end{proposition}
\begin{proof}
    First, we prove that if there exists a homomorphism $\theta$ from $\atoma$ to a fact $\atomb = p(b_1,\dots,b_k) \in I$, then there exists a root-to-leaf path $t$ in $T_p$ s.t. $\theta(\atoma[i]) = t[i]$ $1 \leq i \leq k$.
    By definition of fact homomorphism, $\pred(\atoma) = \pred(\atomb) = p$, for each $\atoma[i] \in \const(\atoma)$, $\theta(\atoma[i]) = \atoma[i] = \atomb[i]$, and for each $\atoma[i] \in \nulls(\atoma)$, $\theta(\atoma[i]) = \atomb[i]$ ($1 \leq i \leq k$).
    Since $\atomb \in I$ then, by definition of af-tree, $T_p$ features a root-to-leaf path $t$ s.t. $t[i] = b[i]$.
    Thus, for each $\atoma[i] \in \const(\atoma)$, $\theta(\atoma[i]) = \atoma[i] = t[i]$, and for each $\atoma[i] \in \nulls(\atoma)$, $\theta(\atoma[i]) = t[i]$, by hypothesis.

    \noindent
    Next, we prove that if there exists $t$ in $T_p$ and a function $\theta$ s.t. $\theta(\atoma[i]) = t[i]$, $1 \leq i \leq k$, then $\theta$ is a homomorphism from $\atoma$ to a fact $\atomb \in I$.
    Indeed, by definition of af-tree, the existence of $T_p$ entails that there exists a fact $\atomb = p(b_1,\dots,b_k) \in I$ represented by a root-to-leaf path $t$ in $T_p$, that is, s.t. $b[i] = t[i]$, $1 \leq i \leq k$.
    Thus, $\theta$ is a homomorphism from $\atoma$ to $\atomb \in I$, as they belong to the same predicate $p$, for each $\atoma[i] \in \const(\atoma)$, $\theta(\atoma[i]) = \atoma[i] = \atomb[i]$, and $\forall \atoma[i] \in \nulls(\atoma)$ $\theta(\atoma[i]) = b[i]$.    
\end{proof}

\subsection{Hash-based Isomorphic Check}

\begin{definition}
Let $\nullsD^c$ be a set of numbered labelled nulls $\{\anull^c_1,\anull^c_2,\dots,\anull^c_n\}$ not appearing in $\nullsD$ or in the set $I$ of facts generated in the chase. 
The \textnormal{canonicalization} of a fact $\atoma$, denoted $\canonical(\atoma)$, is a new fact $\atoma^c$ s.t.: (i)~$\pred(\atoma^c) = \pred(\atoma)$; (ii)~if $\atoma[i] \in \const(\atoma)$, then $\atoma^c[i] = \atoma[i]$; and, (iii)~if $\atoma[i] \in \nulls(\atoma)$, then $\atoma^c[i] = \anull^c_j$, where $j=\arg\min_{k} \atoma[k] = \atoma[i]$. A fact $\atoma$ is \textit{canonical} if $\nulls(\atoma)\subseteq\nullsD^c$.
\end{definition}

\begin{lemma}
Let $\atoma$ be a fact and $\atoma^c$ its $\canonical(\atoma)$.
Then there exists an isomorphism between $\atoma$ and $\atoma^c$.
\end{lemma}
\begin{proof}
    First, we show that there exists a homomorphism $\theta$ from $\atoma$ to $\atoma^c=\canonical(\atoma)$.
    Indeed, by~\Cref{def:null-canonicalization}, they feature the same constants in the same positions, thus $\theta(c)=c$, $\forall c \in \const(\atoma)$.
    Let $\theta^\prime$ be a mapping from $\atoma[i]=\anull_x$ $\in \nulls(\atoma)$ to $\anull^c_j$, where $j$ is the index of the first occurrence of $\anull_x$ in $\atoma$. 
    Thus, $\theta(\anull_x)=\theta'(\anull_x)=\anull^c_j$, which entails that $\theta$ is a homomorphism for all occurrences of $\anull_x$ in $\atoma$ and for each $\anull_x \in \nulls(\atoma)$, since $\atoma^c[i]=\anull^c_j=\theta(\anull_x)=\theta(\atoma[i])$ by~\Cref{def:null-canonicalization}.
    Now, observe that $\theta$ is a bijective function, i.e., for all positions $i$ and $k$ s.t. $\theta(\atoma^c[i])=\theta(\atoma^c[k])$  it holds that $\atoma[i]=\atoma[k]$, as by construction $\theta$ maps to the first occurrence in $\atoma$ of the same null.
\end{proof}

\begin{proposition}
    Let $\atoma$, $\atomb$ be two facts and $\atoma^c$, $\atomb^c$ their canonical form, respectively.
    Then $\atoma$ is isomorphic to $\atomb$ iff $\atoma^c = \atomb^c$.
\end{proposition}
\begin{proof}
    Assume that there exist a isomorphism $\iso$ between $\atoma$ and $\atomb$.
    Clearly, $\pred(\atoma)=\pred(\atoma^c)=\pred({\atomb^c})=\pred({\atomb})$, by definition of fact isomorphism and by~\Cref{def:null-canonicalization}.
    Moreover, $\atoma^c[i] = \atomb^c[i]$, for all positions $i=1,\dots,\arity(\atoma)$.
    Indeed, consider a generic position $i$ of $\atoma^c$ and $\atomb^c$. 
    If $\atoma[i] \in \const(\atoma)$, then $\atoma^c[i] = \atomb^c[i]$, by $\iso$ and by~\Cref{def:null-canonicalization}.
    If $\atoma[i] \in \nulls(\atoma)$ instead, and $\anull^c_j$ is the canonicalization of $\atoma[i]$, then, by~\Cref{def:null-canonicalization}, $j$ is the smallest index s.t. $\atoma[j] = \atoma[i]$. 
    Thus, by definition of $\iso$, $\atomb[j]=\iota(\atoma[j])=\iota(\atoma[i])=\atomb[i]$.
    Finally, by definition of $\canonical$, $\atomb^c[i] = \anull^c_j$, and thus $\atomb^c[i] = \atoma^c[i]$.

    \noindent
    Now, assume that $\atoma^c=\atomb^c$ instead.
    By Lemma~\ref{lemma:canonical-is-isomorphic}, there exists a isomorphism $\iso_\atoma$ between $\atoma$ and $\atoma^c$ and an isomorphism $\iso_\atomb$ between $\atomb$ and $\atomb^c$.
    Consider the composition $\iso = \iso_\atoma \circ \iso_{id} \circ \iso_\atomb^{-1}$, where $\iso_{id}$ is the identity function.
    By definition, $\iso$ is a isomorphism between $\atoma$ and $\atomb$.
\end{proof}

\section{The Chase in Streaming-based Architectures}
First, we recall some definitions and results from the relevant literature.

\begin{mydefinition}[Definition 3.4 (Parsimony) \cite{LeoneMTV19}]
A pair of database $D$ and set of rules $\Sigma$ is called \emph{parsimonious} if, for each atom $\atoma$ of $\ochase(D,\Sigma)$, there exists a homomorphism from $\{\atoma\}$ to $\pchase(D,\Sigma)$.
\end{mydefinition}

\begin{mydefinition}[Definition 4.2 (Nucleus) \cite{LeoneMTV19}]
Consider a set of rules $\Sigma$ and a database $D$. A set $X \subseteq \ochase(D,\Sigma)$ is a \emph{nucleus}
of $\ochase(D,\Sigma)$ if, for each $Y \subseteq \ochase(D,\Sigma)$ containing nulls, there exists a homomorphism
from $Y$ to $\ochase(D,\Sigma)$ that maps at least one null of $Y$ to a term of $X$.    
\end{mydefinition}

\begin{mydefinition}[Definition 4.4 (Strong Parsimony) \cite{LeoneMTV19}]
A pair of database $D$ and set of rules $\Sigma$ is called \emph{strongly parsimonious} if the
following conditions--respectively called \emph{uniformity} and \emph{compactness}-- are satisfied:
\begin{enumerate}
    \item $\langle D \cup F, \Sigma \rangle$ is parsimonious, for each set $F$ of facts; and
    \item $\pchase(D,\Sigma)$ is a nucleus of $\ochase(D,\Sigma)$.
\end{enumerate}
\end{mydefinition}

\begin{myproposition}[Proposition 4.5 \cite{LeoneMTV19}]
    \label{prop:sps-cup-facts-still-sps}
    Consider the strongly parsimonious pair $\langle D, \Sigma\rangle$. For each set $F$ of facts, the pair $\langle D \cup F, \Sigma\rangle$ is still strongly parsimonious.
\end{myproposition}

\begin{mytheorem}[Theorem 5.6 \cite{LeoneMTV19}]
    $\mathit{Shy} \subseteq \mathit{Strongly} \mathit{Parsimonious} \mathit{Sets}$.
\end{mytheorem}

\noindent
We prove the following results under Shy since it is now integrated in Vadalog and Protected is its sub-fragment.
Note that two execution of $\pchaser$, potentially with different rule activation orderings, might yield different sets of generated facts. However, such sets would only differ with respect to the labelled nulls involved in the generated facts~\cite{DBLP:conf/jelia/AmendolaM19}. By definition of Shy, a set $\Sigma$ of rules belonging to such a fragment does not feature attacked harmful joins, that is, no rule features a join between variables in positions invaded by at least one common $\exists$-variable. By property of range disjointness of labelled nulls~\cite{BellomariniSG18}, this entails that no rule is activated on a join between labelled nulls. Therefore, the sets of generated facts are equivalent modulo fact isomorphism, for every database $D$.
Taking this into account, we first prove the following result.

\begin{myproposition}
\label{prop:pchase-modelling}
For any database $D$, a Shy set $\Sigma$ of rules, a BCQ $\query = \psi(\bar{z})$, let $\pchaser^1(D,\Sigma,r)$ and $\pchaser^2(D,\Sigma,r)$ two different executions of $\pchaser$ (i.e. potentially with different rule activation orderings).
We have that $\pchaser^1\\(D,\Sigma,r)\models q$ iff $\pchaser^2(D,\Sigma,r)\models q$.
\end{myproposition}
\begin{proof}
    We prove the claim by induction on the number of resumptions $r$. \\ 
    \noindent
    \emph{Base case.} Let $r=1$. Observe that, in this case, resumption is not involved. 
    If $\pchaser^1(D,\Sigma,1)\models \query$, then by definition there exists a homomorphism $\theta$ from $q$ to $\pchaser^1(D,\Sigma,1)$, i.e. $\theta(\psi(\bar{z})) = \{\theta(\atoma_1),\dots,\theta(\atoma_m)\} \subseteq \pchaser^1(D,\Sigma, 1)$.
    Let $M$ be the number of distinct nulls in the set $\theta(\psi(\bar z))$.
    By the strong parsimony of Shy, and in particular by the definition of nucleus, since $\{\theta(\atoma_1),\dots,\theta(\atoma_n)\}\subseteq \pchaser^1(D,\Sigma,1) \subseteq \ochase(D,\Sigma)$, there exists a homomorphism $\theta_1$ that maps at least one null from $\theta(\psi(\bar z))$ to $\pchaser^2(D,\Sigma,1)$, yielding the set of atoms $\theta_1(\theta(\psi(\bar z)))$.
    Since $\theta_1(\theta(\psi(\bar z))) \subseteq \pchaser^2(D,\Sigma,1)\subseteq \ochase(D,\Sigma)$, again by the definition of nucleus, 
    we can repeat this operation (at most $M-1$ times) for the remaining $M-1$ nulls, with $\theta_2,\dots,\theta_M$ homomorphisms, yielding a set of atoms with no nulls.
    Let $\theta' = \theta \circ \theta_1 \circ \cdots \circ \theta_M$ be the composition of $\theta$ with the succession of the new homomorphism functions.
    By construction, we have that $\theta'$ is a homomorphism from $\psi(\bar z)$ to $\pchaser^2(D,\Sigma,1).$
    The same arguments can be applied in the other direction.
    \\
    \noindent \emph{Inductive case.} Assume the claim holds for all resumption iterations $1,\dots,r-1$, we want to prove that claim holds for $r$.  
    We prove the direction ($\Rightarrow$) of the claim, but the same arguments can be applied in the other direction by swapping $\pchaser^1$ with $\pchaser^2$.
    We distinguish two cases: (i)~$\pchaser^1(D,\Sigma,r-1)\models q$ and (ii)~$\pchaser^1(D,\Sigma,r-1)\not\models q$.
    (i)~If $\pchaser^1(D,\Sigma,r-1)\models q$, we have by inductive hypothesis that $\pchaser^2(D,\Sigma,r-1)\models q$. By monotonicity of the $\pchaser$ operator, we necessarily have that both $\pchaser^1(D,\Sigma,r)\models q$ and $\pchaser^2(D,\Sigma,r)\models q$, since $\pchaser(D,\Sigma,r-1)\subseteq \pchaser(D,\Sigma,r)$.
    (ii)~If $\pchaser^1(D,\Sigma,r-1)\not\models q$, by inductive hypothesis we have that $\pchaser^2(D,\Sigma,r-1)\not\models q$.
    Assume by contradiction that $\pchaser^1(D,\Sigma,r)\models q$ and $\pchaser^2(D,\Sigma,r)\not \models q$. This means that there exists a homomorphism $\theta^1$ from $\psi(\bar z)$ to $\pchaser^1(D,\Sigma, r)$, and that there does not exist a homomorphism from $\psi(\bar z)$ to $\pchaser^2(D,\Sigma,r)$.
    By definition $\pchaser(D,\Sigma,r) = \pchase(D\cup \lceil\pchaser(D,\Sigma,r-1)\rfloor, \Sigma)$.
    Moreover, by Proposition~\ref{prop:sps-cup-facts-still-sps}, we have that the pair $\langle D\cup \lceil\pchaser(D,\Sigma,r-1)\rfloor, \Sigma \rangle$ is strongly parsimonious.
    Therefore, $\pchaser(D,\Sigma,r)$ is a nucleus of $\ochase(D,\Sigma)$;
    so we can apply the same arguments of the base case, and we can always construct a new function $\theta^2$, starting from $\theta^1$, which is a valid homomorphism from $\psi(\bar z)$ to $\pchaser^2(D,\Sigma, r)$.
    Thus $\pchaser^2(D,\Sigma,r)\models q$, which is a contradiction.
\end{proof}

\begin{theorem}
    For any database $D$, a Shy set $\Sigma$ of rules, a BCQ $\query = \psi(\bar{z})$, we have that $\pchase_r(D,\Sigma) \models q$ iff $\chases(D,\Sigma)\models q$ with firing condition $\homomorphismcheckstreaming$.
\end{theorem}
\begin{proof}
    Consider the case when $\mathit{maxRes} = 1$.
    By construction of Algorithm~$3$, all the applicable homomorphisms that satisfy the $\homomorphismcheckstreaming$ firing condition are fired, which means that $\chases(D,\Sigma)$ correctly computes $\pchase(D,\Sigma)$ by \Cref{prop:homomorphic-streaming} and Proposition~\ref{prop:pchase-modelling}.
    Moreover, for a generic $\mathit{maxRes}$, each fact at resumption $r-1$ that is not frozen is frozen and proposed in the resumption iteration $r$, by construction of Algorithm~$3$.
    Again, the correctness follows by \Cref{prop:homomorphic-streaming} and by the fact that the order of rule activations in $\pchase_r$ does not affect the answer to $\query$, as stated in Proposition~\ref{prop:pchase-modelling}.
\end{proof}

\begin{theorem}
    For any database $D$, a Shy set $\Sigma$ of rules, a BCQ $\query = \psi(\bar{z})$, we have that $\ichaser(D,\Sigma) \models q$ iff $\chases(D,\Sigma) \models q$ with firing condition $\isomorphismcheckstreaming$.
\end{theorem}
\begin{proof}
For any resumption iteration, if $\ichaser(D,\Sigma) \models q$, then there exists a homomorphism $\theta$ from $\psi(\bar{z})$ to $\ichaser(D,\Sigma)$. By Lemma~\ref{lemma:pchase-subseteq-ichase}, and by construction of $\pchaser$ there is a homomorphism $\theta'$ that maps $h(\psi(\bar{z}))$ to $\pchaser(D,\Sigma)$. Therefore, $\theta''=\theta\circ \theta'$ is a homomorphism that maps $\psi(\bar{z})$ to $\pchase_r(D,\Sigma)$, i.e. $\pchase_r(D,\Sigma)\models q$. By \Cref{thm:pchase-streaming}, it holds that $\chases(D,\Sigma)\models q$ with firing condition $\homomorphismcheckstreaming$.
Let $\theta_s$ be the homomorphism from $q$ to $\chases(D,\Sigma)$ with firing condition $\homomorphismcheckstreaming$.
By construction, $\chases(D,\Sigma)$ with firing condition $\isomorphismcheckstreaming$ $\supseteq \chases(D,\Sigma)$ with firing condition $\homomorphismcheckstreaming$, and therefore $\theta_s$ is a valid homomorphism from $\psi(\bar{z})$ to $\chases(D,\Sigma)$ with firing condition $\isomorphismcheckstreaming$. The other direction of the claim follows by considering the same arguments swapping $\ichaser$ (resp., $\pchaser$) with $\chases$ employing $\isomorphismcheckstreaming$ (resp., $\homomorphismcheckstreaming$) as firing condition.
\end{proof}

\smallskip
Indeed, by combining the results of Theorem~\ref{thm:pchase-streaming} and Theorem~\ref{thm:bcqa-Shy-pchaser}, we derive the correctness of $\chases$ with firing condition $\homomorphismcheckstreaming$ for BCQA.
Similarly, from Theorem~\ref{thm:ichase-streaming} and Theorem~\ref{thm:bcqa-Shy-ichaser}, we derive the correctness of $\chases$ with firing condition $\isomorphismcheckstreaming$ for BCQA.
\end{document}